\documentclass[useAMS]{article}
\usepackage{graphicx}
\usepackage{natbib} 
\usepackage{amssymb}
\begin{document}

\def\mpch {$h^{-1}$ Mpc} 
\def\kpch {$h^{-1}$ kpc} 
\def\kms {km s$^{-1}$} 
\def\lcdm {$\Lambda$CDM } 
\def\xir {$\xi(r)$}
\def\wprp {$w_p(r_p)$}
\def\xisp {$\xi(r_p,\pi)$}
\def\xis {$\xi(s)$}
\def\rr {$r_0$}
\def\wt {$\omega(\theta)$}
\def\etal {et al.}
\def\deg {$^{\circ}$}
\def\sig {$\sigma_{12}$}

\def\arcmin {$^{\prime}$}
\def\msun{\hbox{$M_{\odot}$}}
\def\hmsun{\hbox{$h^{-1} M_{\odot}$}}

\title{Large Scale Structure of the Universe}
\author{Alison L. Coil \\ 
University of California, San Diego \\ 
La Jolla, CA 92093 \\ 
acoil@ucsd.edu}
\maketitle

\section*{\bf Abstract}
Galaxies are not uniformly distributed in space.  On large scales the
Universe displays coherent structure, with galaxies residing in groups
and clusters on scales of $\sim$1-3 \mpch, which lie at the
intersections of long filaments of galaxies that are $>$10 \mpch \ in
length.  Vast regions of relatively empty space, known as voids, 
contain very few galaxies and span the volume in between these structures. 
This observed large scale structure depends both on cosmological parameters
and on the formation and evolution of galaxies.  
Using the two-point correlation function, one can
trace the dependence of large scale structure on galaxy properties
such as luminosity, color, stellar mass, and track its evolution with
redshift.  Comparison of the observed galaxy clustering signatures
with dark matter simulations allows one to model and understand the
clustering of galaxies and their formation and evolution within their 
parent dark matter halos.
Clustering measurements can determine the parent dark matter halo
mass of a given galaxy population, connect observed galaxy populations
at different epochs, and constrain cosmological parameters and galaxy
evolution models.  This chapter describes the methods used to measure
the two-point correlation function in both redshift and real space,
presents the current results of how the clustering amplitude depends
on various galaxy properties, and discusses quantitative measurements of
the structures of voids and filaments.  The interpretation of these results 
with current theoretical models is also presented.



\section{Historical Background}

Large scale structure is defined as the structure or inhomogeneity 
of the Universe on scales larger than that of a galaxy.  The idea of
whether galaxies are distributed uniformly in space can be traced to
Edwin Hubble, who used his catalog of 400 ``extragalactic nebulae'' to
test the homogeneity of the Universe \citep{Hubble26}, finding it to
be generally uniform on large scales.  In 1932, the larger
Shapley-Ames catalog of bright galaxies was published
\citep{Shapley32}, in which the authors note ``the general unevenness
in distribution'' of the galaxies projected onto the plane of the sky
and the roughly factor of two difference in the numbers of galaxies in
the northern and southern galactic hemispheres.  Using this larger
statistical sample, \citet{Hubble34} noted that on angular scales less
than $\sim$10\deg \ there is an excess in the number counts of galaxies
above what would be expected for a random Poisson distribution, though
the sample follows a Gaussian distribution on larger scales.  Hence,
while the Universe appears to be homogeneous on the largest scales, on
smaller scales it is clearly clumpy.

Measurements of large scale structure took a major leap forward with
the Lick galaxy catalog produced by \citet{Shane67}, which contained
information on roughly a million galaxies obtained using photographic
plates at the 0.5m refractor at Lick Observatory.  \citet{Seldner77}
published maps of the counts of galaxies in angular cells across the
sky (see Fig.~1), which showed in much greater detail that the
projected distribution of galaxies on the plane of the sky is not
uniform.  The maps display a rich structure with a foam-like pattern,
containing possible walls or filaments with long strands of galaxies,
clusters, and large empty regions.  The statistical spatial
distribution of galaxies from this catalog and that of
\citet{Zwicky68} was analyzed by Jim Peebles and collaborators in a
series of papers \citep[e.g.,][]{Peebles75} that showed that the
angular two-point correlation function (defined below) roughly follows
a power law distribution over angular scales of $\sim$0.1\deg --
5\deg.  In these papers it was discovered that the clustering
amplitude is lower for fainter galaxy populations, which likely arises
from larger projection effects along the line of sight.  As faint
galaxies typically lie at larger distances, the projected clustering
integrates over a wider volume of space and therefore dilutes the effect.

\begin{figure}
\begin{center}
  \scalebox{0.7}{\includegraphics{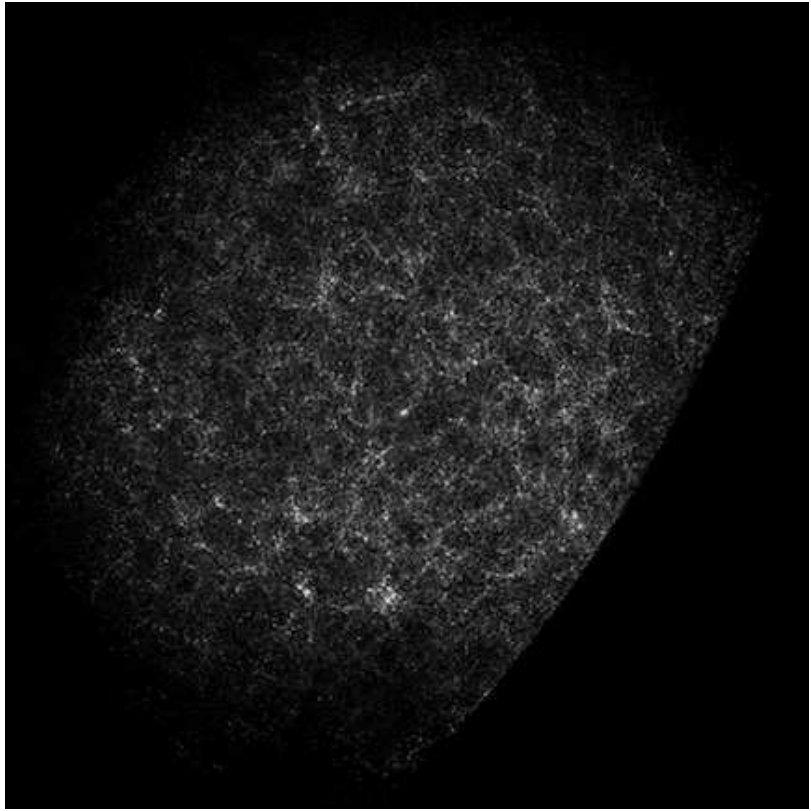}}
\end{center}
\caption{Angular distribution of counts of galaxies brighter than
  $B\sim19$ on the plane of the sky, reconstructed from the Lick
  galaxy catalog (from Seldner \etal\ 1997).  This image shows the
  number of galaxies observed in 10\arcmin \ $\times$ 10\arcmin
  \ cells across the northern galactic hemisphere, where brighter
  cells contain more galaxies. The northern galactic pole is at the
  center, with the galactic equator at the edge.  The distribution of
  galaxies is clearly not uniform; clumps of galaxies are seen in
  white, with very few galaxies observed in the dark regions between.
}
\end{figure}

These results in part spurred the first large scale redshift surveys,
which obtained optical spectra of individual galaxies in order to
measure the redshifts and spatial distributions of large galaxy
samples.  
Pioneering work by \citet{Gregory78} mapped the three-dimensional spatial 
distribution of 238 galaxies around and towards the Coma/Abell 1367 
supercluster.  In addition to surveying the galaxies in the supercluster, 
they found that in the foreground at lower redshift there were large regions 
($>20$ \mpch, shown well in their Figure 2a) with no galaxies, which they 
termed 'voids'.   \citet{Joeveer78} used redshift information 
from \citet{Sandage75} to map the three-dimensional distribution of galaxies 
on large scales in the southern galactic hemisphere.  They mapped four 
separate volumes that included clusters as well as field galaxies and showed 
that across large volumes, galaxies are clearly clustered in three dimensions 
and often form 'chains' of clusters (now recognized as filaments).

Two additional redshifts surveys were the KOS
survey \citep[Kirshner, Oemler, Schechter; ][]{Kirshner78} and the
original CfA survey \citep[Center for Astrophysics; ][]{Davis82dist}.
The KOS survey measured redshifts for 164 galaxies brighter than
magnitude 15 in eight separate fields on the sky, covering a total of
15 deg$^2$.  Part of the motivation for the survey was to study the
three dimensional spatial distribution of galaxies, about which the
authors note that ``although not entirely unexpected, it is striking
how strongly clustered our galaxies are in velocity space,'' as seen
in strongly peaked one dimensional redshift histograms in each field.

The original CfA survey, completed in 1982, contained redshifts for
2,400 galaxies brighter than magnitude 14.5 across the north and south
galactic poles, covering a total of 2.7 steradians.  The major aims of
the survey were cosmological and included quantifying the clustering
of galaxies in three dimensions.  This survey produced large
area, moderately deep three dimensional maps of large scale
structure (see Fig.~2), in which one could identify galaxy clusters, 
voids, and an apparent ``filamentary connected
structure'' between groups of galaxies, which the authors caution
could be random projections of distinct structures
\citep{Davis82dist}.  This paper also performed a comparison of the
so-called ``complex topology'' of the large scale structure seen in
the galaxy distribution with that seen in N-body dark matter
simulations, paving the way for future studies of theoretical models
of structure formation.

The second CfA redshift survey, which ran from 1985 to 1995, contained
spectra for $\sim$5,800 galaxies and revealed the existence of the
so-called ``Great Wall'', a supercluster of galaxies that extends over
170 \mpch, the width of the survey \citep{Geller89}.  Large underdense
voids were also commonly found, with a density 20\% of the mean
density.

\begin{figure}
\begin{center}
\scalebox{0.3}{\includegraphics{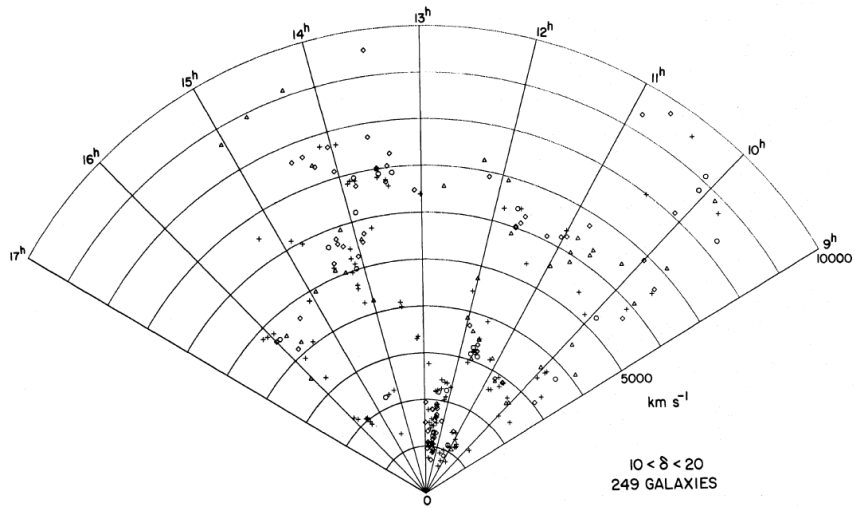}}
\end{center}
\caption{Distribution of galaxies in redshift space from
the original CfA galaxy redshift survey
(from Davis \etal \ 1982).  Plotted are 249 galaxies as a function
of observed velocity (corresponding to a given redshift) 
versus right ascension for a wedge in declination of 
10\deg $< \delta <$ 20\deg. 
}
\end{figure}

Redshift surveys have rapidly progressed with the development of
multi-object spectrographs, which allow simultaneous observations of
hundreds of galaxies, and larger telescopes, which allow deeper surveys of
both lower luminosity nearby galaxies and more distant, luminous galaxies.  
At present the largest redshift surveys of galaxies 
at low redshift are the Two Degree Field 
Galaxy Redshift Survey \citep[2dFGRS,][]{Colless01} and the 
Sloan Digital Sky Survey \citep[SDSS,][]{York00}, which cover volumes of 
$\sim4\times10^7 \ h^3$ Mpc$^{-3}$ and $\sim2\times10^8 \ h^3$ Mpc$^{-3}$ 
with spectroscopic redshifts for $\sim$220,000 and a million galaxies, respectively.
These surveys provide the best current maps of large scale 
structure in the Universe today (see Fig.~3), revealing a sponge-like 
pattern to the distribution of galaxies \citep{Gott86}.  
Voids of $\sim$10 \mpch \ are 
clearly seen, containing very few galaxies.  Filaments stretching 
greater than 10 \mpch \ surround the voids and intersect at the locations 
of galaxy groups and clusters.  

The prevailing theoretical paradigm regarding the existence of large
scale structure is that the initial fluctuations in the energy density
of the early Universe, seen as temperature deviations in the cosmic
microwave background, grow through gravitational instability into the
structure seen today in the galaxy density field.  The details of
large scale structure -- the sizes, densities, and distribution of the
observed structure -- depend both on cosmological parameters such as
the matter density and dark energy as well as on the physics of galaxy
formation and evolution.  Measurements of large scale structure can
therefore constrain both cosmology and galaxy evolution physics.

\begin{figure}
\begin{center}
\scalebox{0.5}{\includegraphics{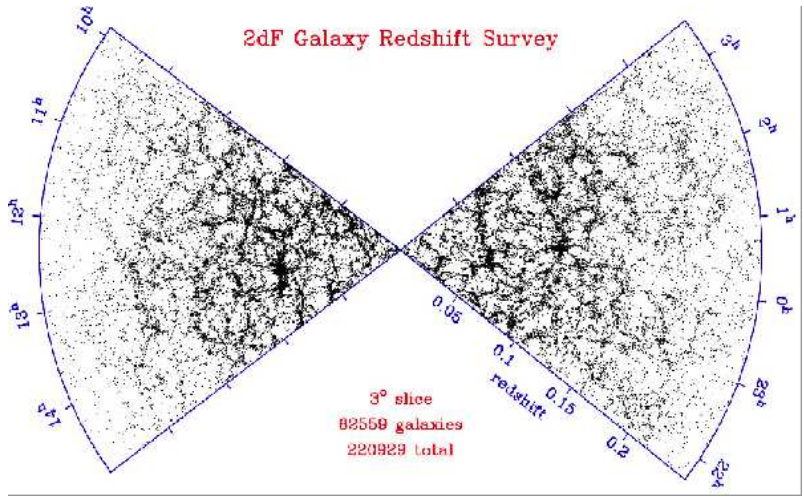}}
\end{center}
\caption{The spatial distribution of galaxies as a function of 
redshift and right ascension (projected through 3\deg \ in declination) 
from the 2dF Galaxy Redshift Survey
(from Colless \etal \ 2004).  
}
\end{figure}

\section{The Two-Point Correlation Function}

In order to quantify the clustering of galaxies, one must survey not
only galaxies in clusters but rather the entire galaxy density
distribution, from voids to superclusters.  The most commonly used
quantitative measure of large scale structure is the galaxy two-point
correlation function, \xir, which traces the amplitude of galaxy
clustering as a function of scale.  \xir \ is defined as a measure of
the excess probability $dP$, above what is expected for an unclustered
random Poisson distribution, of finding a galaxy in a volume element
$dV$ at a separation $r$ from another galaxy,
\begin{equation}
dP = n [1+\xi(r)] dV,
\end{equation}
where $n$ is the mean number density of the galaxy sample in question
\citep{Peebles80}. 
Measurements of \xir \ are generally performed in comoving space, with
$r$ having units of \mpch.
The Fourier transform of the two-point correlation function is the 
power spectrum, which is often used to describe density fluctuations observed 
in the cosmic microwave background.

To measure \xir, one counts pairs of galaxies as a function of
separation and divides by what is expected for an unclustered
distribution.  To do this one must construct a ``random catalog'' that
has the identical three dimensional coverage as the data -- including
the same sky coverage and smoothed redshift distribution -- but is
populated with randomly-distribution points.  The ratio of pairs of
galaxies observed in the data relative to pairs of points in the
random catalog is then used to estimate \xir.  Several different
estimators for \xir \ have been proposed and tested.  An early estimator 
that was widely used is from \citet{Davis83}:
\begin{equation}
\xi = \frac{n_R}{n_D}\frac{DD}{DR}-1,
\end{equation}
where $DD$ and $DR$ are counts of pairs of galaxies (in bins of
separation) in the data catalog and between the data and random catalogs, 
and $n_D$ and $n_R$ are the
mean number densities of galaxies in the data and random catalogs.
\citet{Hamilton93} later introduced an estimator with smaller
statistical errors, 
\begin{equation}
\xi = \frac{DD \ RR}{(DR)^2}-1,
\end{equation}
where $RR$ is the count of pairs of galaxies as a function of 
separation in the random catalog.
The most commonly-used estimator is from \citet{Landy93},
\begin{equation}
\xi=\frac{1}{RR}\left[DD \left(\frac{n_R}{n_D}
\right)^2-2DR\left(\frac{n_R}{n_D} \right)+RR\right].
\end{equation}
This estimator has been shown to perform as well as the Hamilton
estimator (Eqn.~3), and while it requires more computational time 
it is less sensitive to the size of the random catalog and handles edge
corrections well, which can affect clustering measurements on
large scales \citep{Kerscher00}.

As can be seen from the form of the estimators given above, 
measuring \xir \ depends sensitively on having a random
catalog which accurately reflects the various spatial and redshift
selection affects in the data.  These can include effects such as 
edges of slitmasks or fiber plates, overlapping slitmasks or plates, 
gaps between chips on the CCD, and changes in spatial sensitivity 
within the detector (i.e., the effective radial dependence within X-ray
detectors).  If one is measuring a full three-dimensional 
correlation function (discussed below) then the random catalog must
also accurately include the redshift selection of the data. The random
catalog should also be large enough to not introduce Poisson error in
the estimator.  This can be checked by ensuring that the RR pair counts
in the smallest bin are high enough such that Poisson errors are subdominant.

\section{Angular Clustering}

The spatial distribution of galaxies can be measured either in two
dimensions as projected onto the plane of the sky or in three
dimensions using the redshift of each galaxy.  As it can be
observationally expensive to obtain spectra for large samples of
(particularly faint) galaxies, redshift information is not always
available for a given sample.  One can then measure the
two-dimensional projected angular correlation function \wt, defined as
the probability above Poisson of finding two galaxies with an angular
separation $\theta$:
\begin{equation}
dP = N [1+\omega(\theta)] d\Omega
\end{equation}
where $N$ is the mean number of galaxies per steradian and d$\Omega$
is the solid angle of a second galaxy at a separation $\theta$ from a
randomly chosen galaxy.

Measurements of \wt \ are known to be low by an additive factor known
as the ``integral constraint'', which results from using the data
sample itself (which often does not cover large areas of the sky) to
estimate the mean galaxy density.  This correction becomes important
on angular scales comparable to the survey size; on much smaller
scales it is negligible.  One can either restrict measurements of the
angular clustering to scales where the integral constraint is not
important or estimate the amplitude of the integral constraint
correction by doubly integrating an assumed power law form of \wt \ 
over the survey area, $\Omega$, using
\begin{equation}
AC = \frac{1}{\Omega}\int\int \omega(\theta) d\Omega_1 d\Omega_2,
\end{equation}
where $\Omega$ is the area subtended by the survey.  
In practice, this can be numerically estimated over the survey geometry using 
the random catalog itself 
(see Roche \& Eales 1999 for details)\nocite{Roche99}.

The projected angular two-point correlation function, \wt, can generally
be fit with a power law,
\begin{equation}
\omega(\theta)=A_{\omega}\theta^{\delta}
\end{equation}
where A is the clustering amplitude at a given scale (often 1\arcmin) 
and $\delta$ is the slope of the correlation function.  

From measurements of \wt \ one can infer the three-dimensional spatial
two-point correlation function, $\xi(r)$, if the redshift distribution
of the sources is well known.  The two-point correlation function,
$\xi(r)$, is usually fit as a power law, $\xi(r)=(r/r_0)^{-\gamma}$,
where $r_0$ is the characteristic scale-length of the galaxy
clustering, defined as the scale at which $\xi=1$.  As the
two-dimensional galaxy clustering seen in the plane of the sky is a
projection of the three-dimensional clustering, \wt \ is directly
related to its three-dimensional analog $\xi(r)$.  For a given
$\xi(r)$, one can predict the amplitude and slope of \wt \ using
Limber's equation, effectively integrating $\xi(r)$ along the redshift
direction (e.g. Peebles 1980)\nocite{Peebles80}.  If one assumes
$\xi(r)$ (and thus \wt) to be a power law over the redshift range of
interest, such that
\begin{equation}
\xi(r,z)=\left[\frac{r_0(z)}{r} \right]^\gamma,
\end{equation}
then
\begin{equation}
w(\theta)=\sqrt{\pi} \frac{\Gamma[(\gamma-1)/2]}{\Gamma(\gamma/2)}\frac{A}{\theta^{\gamma-1}},
\label{wthetaeqn}
\end{equation}
where $\Gamma$ is the usual gamma function.  
The amplitude factor, $A$, is given by
\begin{equation}
A=\frac{\int_{0}^{\infty} r_0^\gamma(z)
g(z)\left(\frac{dN}{dz}\right)^2
dz}{\left[\int_{0}^{\infty}\left(\frac{dN}{dz} \right) dz \right]^2}
\end{equation}
where $dN/dz$ is the number of
galaxies per unit redshift interval and $g(z)$ depends on $\gamma$ 
and the cosmological model:
\begin{equation}
g(z)=\left(\frac{dz}{dr} \right)r^{(1-\gamma)}F(r).
\label{epseqn}
\end{equation}
Here $F(r)$ is the curvature factor in the Robertson-Walker metric,
\begin{equation}
ds^2=c^2 dt^2 - a^2[dr^2/F(r)^2+r^2 d\theta^2+r^2 sin^2 \theta d\phi^2].
\end{equation}   
If the redshift distribution of sources, $dN/dz$, is well known, then
the amplitude of \wt \ can be predicted for a given power law model of
\xir, such that measurements of \wt \ can be used to place
constraints on the evolution of \xir.

Interpreting angular clustering results can be difficult, however, as
there is a degeneracy between the inherent clustering amplitude and
the redshift distribution of the sources in the sample.  For example,
an observed weakly clustered signal projected on the plane of the sky
could be due either to the galaxy population being intrinsically
weakly clustered and projected over a relatively short distance along
the line of sight, or it could result from an inherently strongly
clustered distribution integrated over a long distance, which would
wash out the signal.  The uncertainty on the redshift distribution is
therefore often the dominant error in analyses using angular
clustering measurements.  The assumed galaxy redshift distribution
($dN/dz$) has varied widely in different studies, such that similar
observed angular clustering results have led to widely different
conclusions.  A further complication is that each sample usually spans
a large range of redshifts and is magnitude-limited, such that the
mean intrinsic luminosity of the galaxies is changing with redshift
within a sample, which can hinder interpretation of the evolution of
clustering measured in \wt \ studies.

Many of the first measurements of large scale structure were studies
of angular clustering. One of the earliest determinations was the
pioneering work of \citet{Peebles75} using photographic plates from the
Lick survey (Fig.~1).  They found \wt \ to be well fit by a power law
with a slope of $\delta=-0.8$.  Later studies using CCDs were able to
reach deeper magnitude limits and found that fainter galaxies had a
lower clustering amplitude.  One such study was conducted by
\citet{Postman98}, who surveyed a contiguous 4\deg \ by 4\deg \ field
to a depth of $I_{\rm AB}=24$ mag, reaching to $z\sim1$.  Later
surveys that covered multiple fields on the sky found similar results.
The lower clustering amplitude observed for galaxies with fainter
apparent magnitudes can in principle be due either to clustering
being a function of luminosity and/or a function of redshift.  To
disentangle this dependence, each author assumes a $dN/dz$
distribution for galaxies as a function of apparent magnitude and then
fits the observed \wt \ with different models of the redshift
dependence of clustering.  While many authors measure similar values
of the dependence of \wt \ on apparent magnitude, due to differences
in the assumed $dN/dz$ distributions different conclusions are
reached regarding the amount of luminosity and redshift dependence to
galaxy clustering.  Additionally, quoted error bars on the inferred
values of \rr \ generally include only Poisson and/or cosmic variance
error estimates, while the dominant error is often the lack of
knowledge of $dN/dz$ for the particular sample in question.

Because of the sensitivity of the inferred value of \rr \ on the
redshift distribution of sources, it is preferable to measure the
three dimensional correlation function.  While it is much easier to
interpret three dimensional clustering measurements, in cases where it
is still not feasible to obtain redshifts for a large fraction of
galaxies in the sample, angular clustering measurements are still
employed.  This is currently the case in particular with high redshift
and/or very dusty, optically-obscured galaxy samples, such as
sub-millimeter galaxies \citep[e.g.,][]{Brodwin08, Maddox10}. However,
without knowledge of the redshift distribution of the sources, these
measurements are hard to interpret.

\section{Real and Redshift Space Clustering}

Measurements of the two-point correlation function use the redshift of
a galaxy, not its distance, to infer its location along the line of
sight.  This introduces two complications: one is that a cosmological
model has to be assumed to convert measured redshifts to inferred
distances, and the other is that peculiar velocities introduce
redshift space distortions in $\xi$ parallel to the line of sight
\citet{Sargent77}.  On the first point, errors on the assumed
cosmology are generally subdominant, so that while in theory one could
assume different cosmological parameters and check which results are
consistent with the assumed values, that is generally not necessary.
On the second point, redshift space distortions can be measured to
constrain cosmological parameters, and they can also be integrated
over to recover the underlying real space correlation function.

On small spatial scales ($\lesssim 1$ \mpch), within collapsed
virialized overdensities such as groups and clusters, galaxies have
large random motions relative to each other.  Therefore while all of
the galaxies in the group or cluster have a similar physical distance
from the observer, they have somewhat different redshifts.  This
causes an elongation in redshift space maps along the line of sight
within overdense regions, which is referred to as ``Fingers of God''.
The result is that groups and clusters appear to be radially extended
along the line of sight towards the observer. This effect can be seen
clearly in Fig.~4, where the lower left panel shows galaxies in
redshift space with large ``Fingers of God'' pointing back to the
observer, while in the lower right panel the ``Fingers of God'' have
been modeled and removed.  Redshift space distortions are also seen on
larger scales ($\gtrsim 1$ \mpch) due to streaming motions of galaxies
that are infalling onto structures that are still collapsing.
Adjacent galaxies will all be moving in the same direction, which
leads to coherent motion and causes an apparent {\it contraction} of
structure along the line of sight in redshift space \citep{Kaiser87},
in the opposite sense as the ``Fingers of God''.

\begin{figure}
\begin{center}
\scalebox{0.6}{\includegraphics{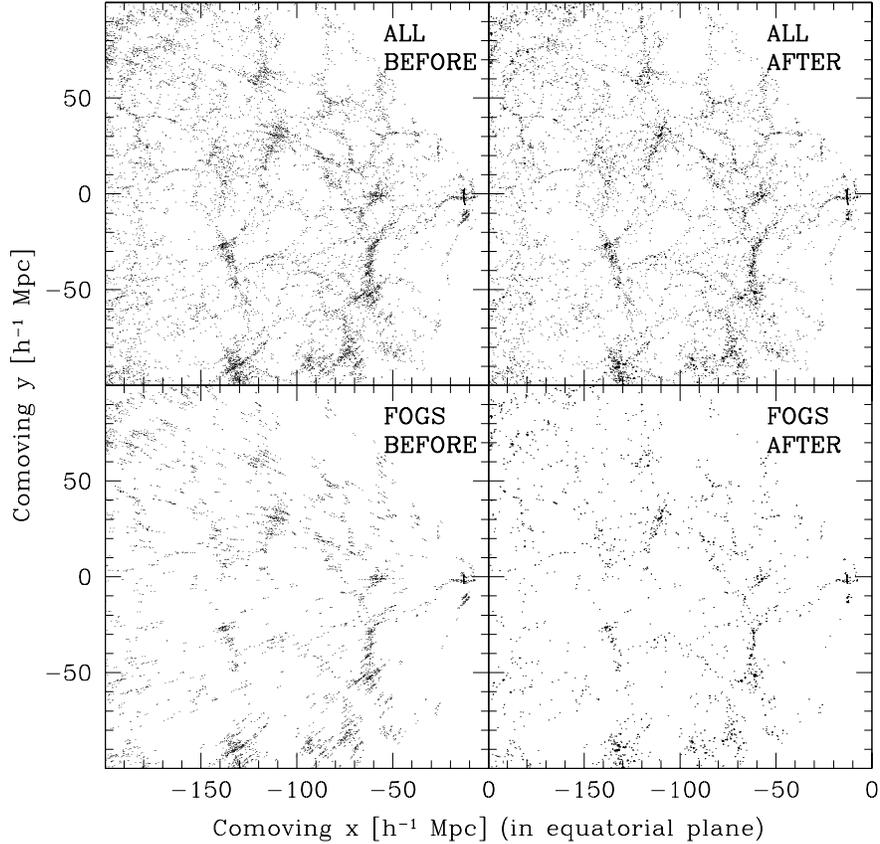}}
\end{center}
\caption{An illustration of the ``Fingers of God'' (FoG), or elongation 
of virialized structures along the line of sight, from \citet{Tegmark04}.
Shown are galaxies from a slice of the SDSS sample (projected here through
the declination direction) in two dimensional comoving space.  The top row
shows all galaxies in this slice (67,626 galaxies in total), while the 
bottom row shows galaxies that have been identified as having ``Fingers of God''.  
The right column shows the position of these galaxies in this space after modeling 
and removing the effects of the ``Fingers of God''.  The observer is located 
at (x,y=0,0), and the ``Fingers of God'' effect can be seen in the lower left 
panel as the positions of galaxies being radially smeared along the line of 
sight toward the observer.
}
\end{figure}

Redshift space distortions can be clearly seen in measurements of
galaxy clustering.  While redshift space distortions can be used to
uncover information about the underlying matter density and thermal
motions of the galaxies (discussed below), they complicate a
measurement of the two-point correlation function in real space.
Instead of \xir, what is measured is \xis, where $s$ is the redshift
space separation between a pair of galaxies.  While some results in
the literature present measurements of \xis \ for various galaxy
populations, it is not straightforward to compare results for
different galaxy samples and different redshifts, as the amplitude of
redshift space distortions differs depending on the galaxy type and
redshift. Additionally, \xis \ does not follow a power law over the
same scales as \xir, as redshift space distortions on both small and
large scales decrease the amplitude of clustering relative to
intermediate scales.

The real-space correlation function, \xir, measures the underlying
physical clustering of galaxies, independent of any peculiar
velocities.  Therefore, in order to recover the real-space correlation
function, one can measure $\xi$ in two dimensions, both perpendicular
to and along the line of sight.  Following \citet{Fisher94}, ${\bf
  v_1}$ and ${\bf v_2}$ are defined to be the redshift positions of a
pair of galaxies, ${\bf s}$ to be the redshift space separation (${\bf
  v_1}-{\bf v_2}$), and ${\bf l} =\frac{1}{2}$(${\bf v_1}+{\bf v_2})$
to be the mean distance to the pair.  The separation between the two
galaxies across ($r_p$) and along ($\pi$) the line of sight are
defined as
\begin{equation}
\pi=\frac{{\bf s} \cdot {\bf l}}{{\bf |l|}},
\end{equation}
\begin{equation}
r_p=\sqrt{{\bf s} \cdot {\bf s} - \pi^2}.
\end{equation}
One can then compute pair
counts over a two-dimensional grid of separations to estimate \xisp.
\xis, the one-dimensional redshift space correlation 
function, is then equivalent to the azimuthal average of \xisp. 

\begin{figure}
\begin{center}
\scalebox{0.6}{\includegraphics{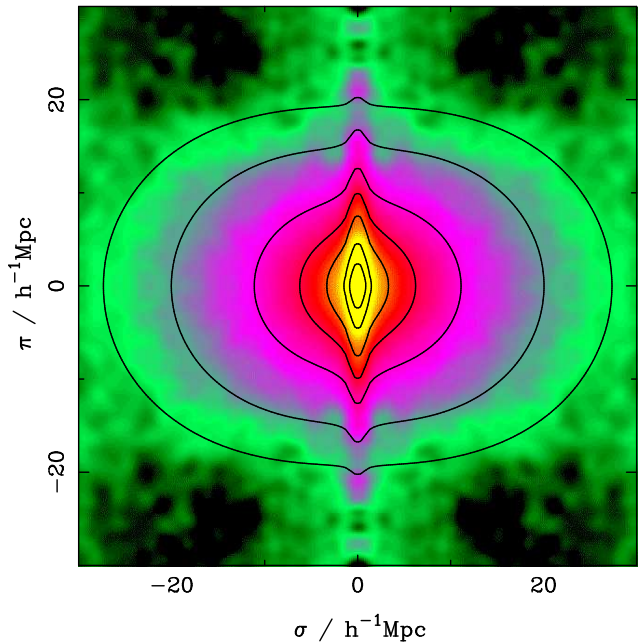}}
\end{center}
\caption{The two-dimensional redshift space correlation function 
from 2dFGRS \citep{Peacock01}.  Shown is \xisp \ (in the figure $\sigma$ is
used instead of $r_p$), the correlation function as a function of
separation across ($\sigma$ or $r_p$) and along ($\pi$) the line of
sight.  Contours show lines of constant $\xi$ 
at $\xi=$10, 5, 2, 1, 0.5, 0.2, 0.1.
Data from the first quadrant (upper right) are reflected about both the $\sigma$ 
and $\pi$ axes, to
emphasize deviations from circular symmetry due to redshift space distortions.
}
\end{figure}

An example of a measurement of \xisp \ is shown in Fig.~5.  Plotted is
$\xi$ as a function of separation $r_p$ (defined in this figure to be
$\sigma$) across and $\pi$ along the line of sight.  What is usually
shown is the upper right quadrant of this figure, which here has been
reflected about both axes to emphasize the distortions.  Contours of
constant $\xi$ follow the color-coding, where yellow corresponds to
large $\xi$ values and green to low values.  On small scales across
the line of sight ($r_p$ or $\sigma < \sim2$ \mpch) the contours are
clearly elongated in the $\pi$ direction; this reflects the ``Fingers
of God'' from galaxies in virialized overdensities.  On large scales
across the line of sight ($r_P$ or $\sigma >\sim10$ \mpch) the
contours are flattened along the line of sight, due to ``the Kaiser
effect''.  This indicates that galaxies on these linear scales are
coherently streaming onto structures that are still collapsing.

As this effect is due to the gravitational infall of galaxies onto
massive forming structures, the strength of the signature depends 
on $\Omega_{\rm matter}$.  
\citet{Kaiser87} derived that the large-scale anisotropy in
the \xisp \ plane depends on 
$\beta \equiv \Omega_{\rm matter}/b$ on linear scales, where $b$ is the bias or 
the ratio of density fluctuations in the galaxy population relative to 
that of dark matter (discussed further in the next section below).
Anisotropies are quantified using the
multipole moments of $\xi(r_p,\pi)$, defined as
\begin{equation}
\xi_l(s) = (2l+1)/2 \int{\xi(r_p,\pi) \ P_l(\cos\theta) \ d\cos\theta},
\end{equation}
where $s$ is the distance as measured in redshift space, $P_l$ are
Legendre polynomials, and $\theta$ is the angle between $s$ and
the line of sight.
The ratio $\xi_2$/$\xi_0$, the quadrupole to monopole moments of the
two-point correlation function, is related to $\beta$ in a simple
manner using linear theory \citep{Hamilton98}:
\begin{equation}
\xi_2/\xi_0=f(n)
\frac{\frac{4}{3}\beta+\frac{4}{7}\beta^2}{1+\frac{2}{3}\beta+\frac{1}{5}\beta^2}
\label{eqnlinear}
\end{equation}
where $f(n)$ =(3+$n)/n$ and $n$ is the index of the two-point
correlation function in a power-law form: $\xi\propto r^{-(3+n)}$
\citep{Hamilton92}.

\citet{Peacock01} find using measurements of the
quadrupole-to-monopole ratio in the 2dFGRS data (see Fig.~5) that
$\beta = 0.43 \pm0.07$.  For a bias value of around unity (see Section
5 below), this implies a low value of $\Omega_{\rm matter} \sim 0.3$.
Similar measurements have been made with clustering measurements using
data from the SDSS.  Very large galaxy samples are needed to detect
this coherent infall and obtain robust estimates of $\beta$.  At
higher redshift, \citet{Guzzo08} find $\beta=0.70 \pm0.26$ at $z=0.77$
using data from the VVDS and argue that measurements of $\beta$ as a
function of redshift can be used to trace the expansion history of the
Universe.  We return to the discussion of redshift space distortions
on small scales below in Section 6.3.

What is often desired, however, is a measurement of the real space 
clustering of galaxies. To recover \xir \ one can then project \xisp \  
along the $r_p$ axis.  As redshift space distortions affect only the
line-of-sight component of \xisp, integrating over the $\pi$ direction
leads to a statistic \wprp, which is independent of redshift space
distortions.  Following \citet{Davis83},
\begin{equation}
w_p(r_p)=2 \int_{0}^{\infty} d\pi \ \xi(r_p,\pi)=2 \int_{0}^{\infty}
dy \ \xi(r_p^2+y^2)^{1/2},
\label{eqn}
\end{equation}
where $y$ is the real-space separation along the line of sight. If
\xir \ is modeled as a power-law, $\xi(r)=(r/r_0)^{-\gamma}$, then \rr \ 
and $\gamma$ can be readily extracted from the projected correlation
function, \wprp, using an analytic solution to Equation \ref{eqn}:
\begin{equation}
w_p(r_p)=r_p \left(\frac{r_0}{r_p}\right)^\gamma
\frac{\Gamma(\frac{1}{2})\Gamma(\frac{\gamma-1}{2})}{\Gamma(\frac{\gamma}{2})},
\label{powerlawwprp}
\end{equation}
where $\Gamma$ is the usual gamma function.  A power-law fit to \wprp \ 
will then recover \rr \ and $\gamma$ for the real-space correlation 
function, \xir.  In practice, Equation \ref{eqn} is not integrated to
infinite separations.  Often values of $\pi_{\rm max}$ are $\sim$40-80
\mpch, which includes most correlated pairs.  It is worth noting that
the values of \rr \ and $\gamma$ inferred are covariant.  One must
therefore be careful when comparing clustering amplitudes of different
galaxy populations; simply comparing the \rr \ values may be
misleading if the correlation function slopes are different.  It is
often preferred to compare the galaxy bias instead (see next section).

As a final note on measuring the two-point correlation function, as can 
be seen from Fig.~3, flux-limited galaxy samples contain a higher density 
of galaxies at lower redshift.  This is purely an observational artifact, 
due to the apparent magnitude limit including intrinsically lower luminosity
galaxies nearby, while only tracing the higher luminosity galaxies further 
away.  As discussed below in Section 6, because the clustering amplitude of
galaxies depends on their properties, including luminosity, one would 
ideally only measure \xir \ in volume-limited samples, where galaxies of
the same absolute magnitude are observed throughout the entire volume 
of the sample, including at the highest redshifts.  Therefore often the
full observed galaxy population is not used in measurements of \xir, rather
volume-limited sub-samples are created where all galaxies are brighter than
a given absolute magnitude limit.  This greatly facilitates the theoretical 
interpretation of clustering measurements (see Section 8) and the comparison 
of results from different surveys.

\section{Galaxy Bias}

It was realized decades ago that the spatial clustering of observable
galaxies need not precisely mirror the clustering of the bulk of the
matter in the Universe.  In its most general form, the galaxy density
can be a non-local and stochastic function of the underlying dark
matter density. This galaxy ``bias'' -- the relationship between the
spatial distribution of galaxies and the underlying dark matter
density field -- is a result of the varied physics of galaxy formation
which can cause the spatial distribution of baryons to differ from
that of dark matter.  Stochasticity appears to have little effect on
bias except for adding extra variance \citep[e.g.,][]{Scoccimarro00},
and non-locality can be taken into account to first order by using
smoothed densities over larger scales. In this approximation, the
smoothed galaxy density contrast is a general function of the
underlying dark matter density contrast on some scale:
\begin{equation}
\delta_g = f(\delta),
\end{equation}
where $\delta \equiv (\rho / \bar{\rho}) -1$ and 
$\bar{\rho}$ is the mean mass density on that scale.  
If we assume $f(\delta)$ is a linear function of $\delta$, then 
we can define the linear galaxies bias $b$ 
as the ratio of the mean
overdensity of galaxies to the mean overdensity of mass,
\begin{equation}
b=\delta_g/\delta, 
\end{equation} 
and can in theory depend on scale and galaxy properties such as 
luminosity, morphology, color and redshift.  
In terms of the correlation function, the linear bias is defined
as the square root of the ratio of the two-point correlation
function of the galaxies relative to the dark matter:
\begin{equation}
b=(\xi_{\rm gal}/\xi_{\rm dark \ matter})^{1/2} 
\end{equation}
and is a function of scale.
Note that $\xi_{\rm dark \ matter}$ is the Fourier transform of the 
dark matter power spectrum.  
The bias of galaxies relative to dark matter is often referred to as 
the absolute bias, as opposed to the relative bias between galaxy populations 
(discussed below).

The concept of galaxies being a biased tracer of the underlying total
mass field (which is dominated by dark matter) was introduced by
\citet{Kaiser84} in an attempt to reconcile the different clustering
scale lengths of galaxies and rich clusters, which could not both be
unbiased tracers of mass.  \citet{Kaiser84} show that clusters of
galaxies would naturally have a large bias as a result of being rare
objects which formed at the highest density peaks of the mass
distribution, above some critical threshold.  This idea is further
developed analytically by \citet{Bardeen86} for galaxies, who show 
that for a Gaussian distribution of initial mass density fluctuations,
the peaks which first collapse to form galaxies will be more clustered
than the underlying mass distribution. \citet{Mo96} use extended
Press-Schechter theory to determine that the bias depends on the mass of the
dark matter halo as well as the epoch of galaxy formation and that a
linear bias is a decent approximation well into the non-linear regime
where $\delta >$1.  The evolution of bias with redshift is developed
in theoretical work by \citet{Fry96} and \citet{Tegmark98}, who find 
that the bias is naturally larger at earlier epochs of galaxy formation,
as the first galaxies to form will collapse in the most overdense
regions of space, which are biased (akin to mountain peaks being
clustered).  They further show that regardless of the initial
amplitude of the bias factor, with time galaxies will become unbiased
tracers of the mass distribution ($b\rightarrow$1 as $t\rightarrow
\infty$).  Additionally, \citet{Mann98} find that while bias is generally
scale-dependent, the dependence is weak and on large scales the bias
tends towards a constant value.

A galaxy population can be ``anti-biased'' if $b<1$, indicating that
galaxies are less clustered than the dark matter distribution.  As
discussed below, this appears to be the case for some galaxy samples
at low redshift.  The galaxy bias of a given observational sample is
often inferred by comparing the observed clustering of galaxies with
the clustering of dark matter measured in a cosmological simulation.
Therefore the bias depends on the cosmological model used in the
simulation.  The dominant relevant cosmological parameter is
$\sigma_8$, defined as the standard deviation of galaxy count
fluctuations in a sphere of radius 8 \mpch, and the absolute bias
value inferred can be simply scaled with the assumed value of
$\sigma_8$.  As discussed in section 9.1 below, the absolute galaxy
bias can also be estimated from the data directly, without having to
resort to comparisons with cosmological simulations, by using the
ratio of the two-point and three-point correlation functions, which
have different dependencies on the bias.  While this measurement can
be somewhat noisy, it has the advantage of not assuming a cosmological
model from which to derive the dark matter clustering.  This
measurement is performed by \citet{Verde02} and \citet{Gaztanaga05},
who find that galaxies in 2dFGRS have a linear bias value very close
to unity on large scales.

The relative bias between different galaxy populations can also be measured
and is defined as the 
ratio of the clustering of one population relative to another.  This is
often measured using the ratio of the projected correlation functions of
each population:
\begin{equation}
b_{\rm gal1/gal2}=(w_{p,\rm gal1}/w_{p,\rm gal2})^{1/2},
\end{equation}
where both measurements of \wprp \ have been integrated to the same
value of $\pi_{max}$.  The relative bias is used to compare the clustering 
of galaxies as a function of observed parameters and does not refer to the 
clustering of dark matter.  It is a useful way to compare the observed
clustering for different galaxy populations without having to rely on an 
assumed value of $\sigma_8$ for the dark matter.

\section{The Dependence of Clustering on Galaxy Properties}

The two-point correlation function has long been known to depend on
galaxy properties and can vary as a function of galaxy luminosity, 
morphological or spectral type, color, stellar mass, and redshift.  
The general trend is that galaxies that are more luminous, early-type,
bulge-dominated, optically red, and/or higher stellar mass are more 
clustered than galaxies that are less luminous, late-type, disk-dominated,
optically blue, and/or lower stellar mass.  Presented below are relatively
recent results indicating how clustering properties depend on galaxy 
properties from the largest redshifts surveys currently available.
The physical interpretation behind these trends is presented in Section 8 below.

\subsection{Luminosity Dependence}

Fig.~3 shows the large scale structure reflected in the galaxy
distribution at low redshift.  What is plotted is the spatial
distribution of galaxies in a flux-limited sample, meaning that all
galaxies down to a given apparent magnitude limit are included.  This
results in the apparent lack of galaxies or structure at higher
redshift in the figure, as at large distances only the most luminous
galaxies will be included in a flux-limited sample.  In order to
robustly determine the underlying clustering, one should, if possible,
create volume-limited subsamples in which galaxies of the same
luminosity can be detected at all redshifts.  In this way the mean
luminosity of the sample does not change with redshift and galaxies at
all redshifts are weighted equally.

The left panel of Fig.~6 shows the projected correlation function,
\wprp, for galaxies in SDSS in volume-limited subsamples corresponding
to different absolute magnitude ranges.  The more luminous galaxies
are more strongly clustered across a wide range in absolute optical
magnitude, from $-17<M_r<-23$.  Power law fits on scales from
$\sim$0.1 \mpch \ to $\sim$10 \mpch \ show that while the clustering
amplitude depends sensitively on luminosity, the slope does not.  Only
in the brightest and faintest magnitude bins does the slope deviate
from $\gamma\sim1.8$ and have a steeper value of $\gamma\sim2.0$.
Across this magnitude range \rr \ varies from $\sim$2.8 \mpch \ at the
faint end to $\sim$10 \mpch \ at the bright end.  This same general
trend is found in the 2dFGRS and other redshift surveys
\citep[e.g.,][]{Norberg01}.

The right panel of Fig.~6 shows the relative bias of SDSS galaxies as
a function of luminosity, relative to the clustering of $L^*$
galaxies, measured at the scale of $r_p=2.7$ \mpch, which is in the
non-linear regime where $\delta>1$ \citep{Zehavi05}.  $L^*$ is the
characteristic galaxy luminosity, defined as the luminosity of the
break in the galaxy luminosity function. The relative bias is seen to
steadily increase at higher luminosity and rise sharply above $L^*$.
This is in good agreement with the results from \citet{Tegmark04},
using the power spectrum of SDSS galaxies measured in the linear
regime on a scale of $\sim100$ \mpch.  The data also agree with the
clustering results of galaxies in the 2dFGRS from \citet{Norberg01}.
The overall shape of the relative bias with luminosity indicates a
slow rise up to the value at $L^*$, above which the rise is much
steeper.  As discussed in Section 8.2 below, this trend shows that
brighter galaxies reside in more massive dark matter halos than
fainter galaxies.

\begin{figure}
\begin{center}
\includegraphics[width=0.45\linewidth]{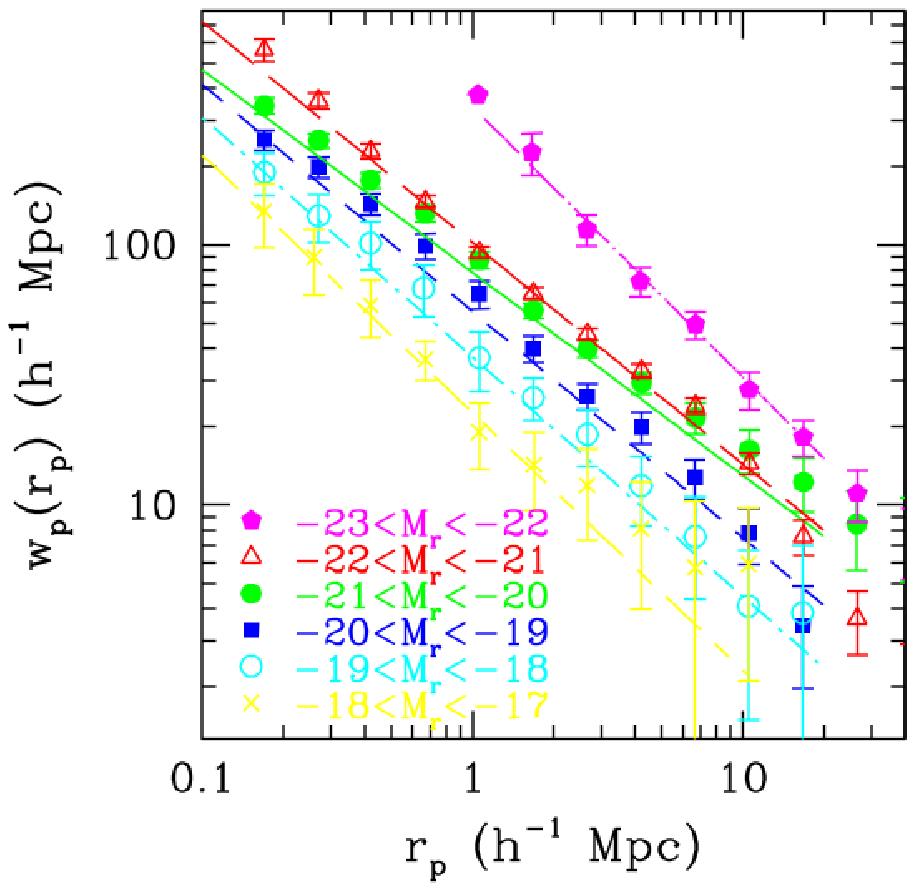}
\includegraphics[width=0.5\linewidth]{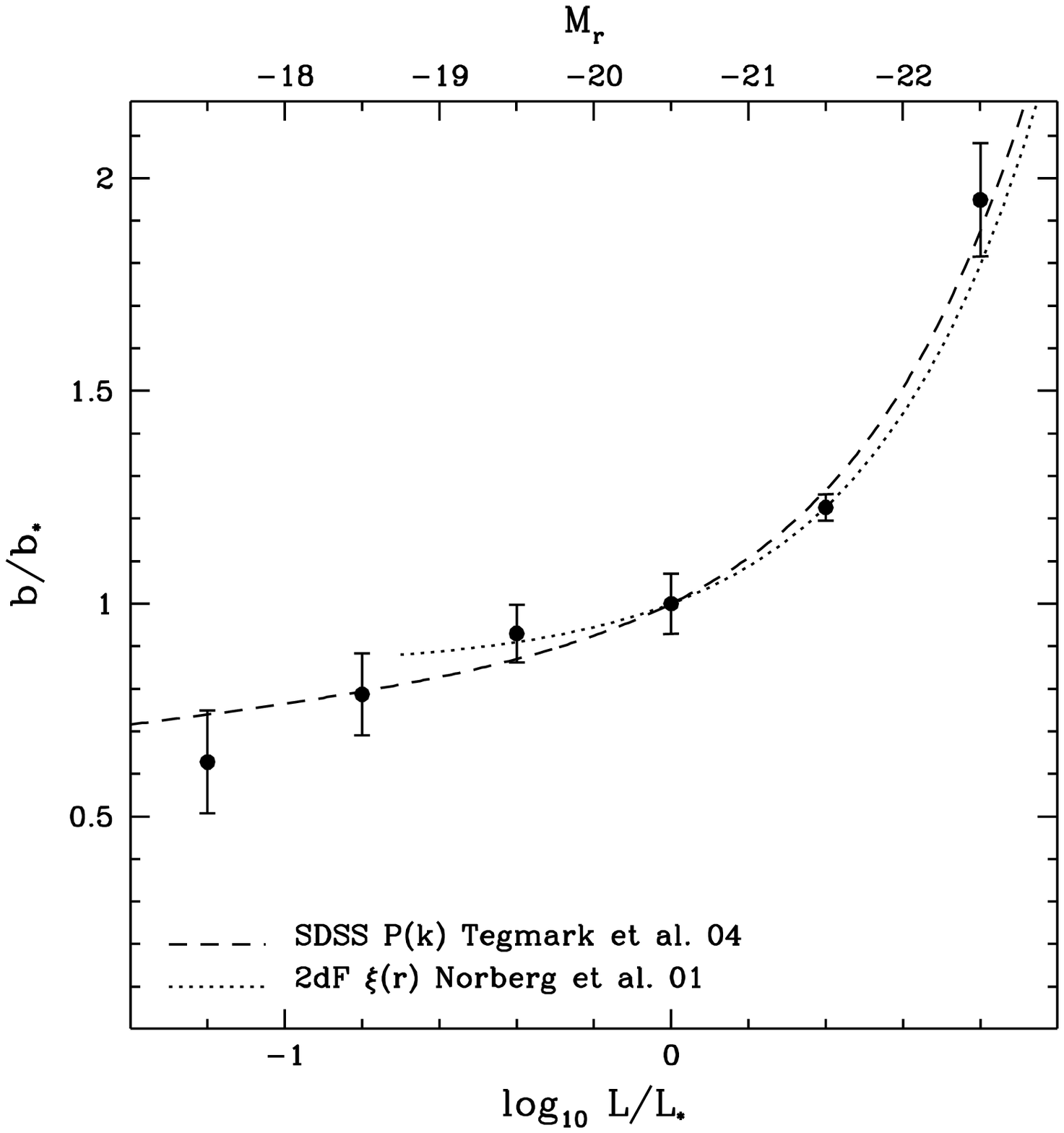}
\end{center}
\caption{Luminosity-dependence of galaxy clustering.  On the left is shown
the projected correlation function, \wprp, for SDSS galaxies in different
absolute magnitude ranges, where brighter galaxies are seen to be more
clustered.  On the right is the relative bias of galaxies as a function of
luminosity.  Both figures are from \citet{Zehavi05}.
}
\end{figure}

\subsection{Color and Spectral Type Dependence}

The clustering strength of galaxies also depends on restframe color
and spectral type, with a stronger dependence than on luminosity.
Fig.~7 shows the spatial distribution of galaxies in SDSS, color coded
as a function of restframe color.  Red galaxies are seen to
preferentially populate the most overdense regions, while blue
galaxies are more smoothly distributed in space.  This is reflected in
the correlation function of galaxies split by restframe color.  Red
galaxies have a larger correlation length and steeper slope than blue
galaxies: \rr$\sim$5-6 \mpch \ and $\gamma\sim$2.0 for red $L^*$
galaxies, while \rr$\sim$3-4 \mpch \ and $\gamma\sim$1.7 for blue
$L^*$ galaxies in SDSS \citep{Zehavi05}.  Clustering studies from the
2dFGRS split the galaxy sample at low redshift by spectral type into
galaxies with emission line spectra versus absorption line spectra,
corresponding to star forming and quiescent galaxies, and find similar
results: that quiescent galaxies have larger correlation lengths and
steeper clustering slopes than star forming galaxies
\citep{Madgwick03}.

Red and blue galaxies have distinct luminosity-dependent clustering
properties.  As shown in Fig.~8, the general trends seen in \rr \ and
$\gamma$ with luminosity for all galaxies are well-reflected in the
blue galaxy population; however, at faint luminosities ($L\lesssim0.5
L^*$) red galaxies have larger clustering amplitudes and slopes than
$L^*$ red galaxies.  This reflects the fact that faint red galaxies
are often found distributed throughout galaxy clusters.

\begin{figure}
\begin{center}
\scalebox{0.6}{\includegraphics{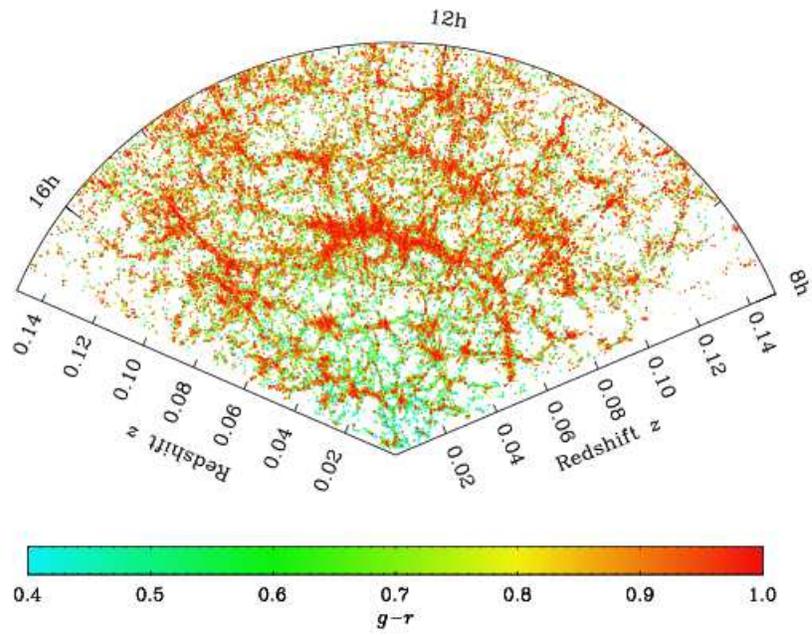}}
\end{center}
\caption{The spatial distribution of galaxies in the SDSS main galaxy sample as 
a function of redshift and right ascension, 
projected through 8\deg \ in declination, 
color coded by restframe optical color.  Red galaxies are seen to be more clustered
than blue galaxies and generally trace the centers of groups and clusters, while
blue galaxies populate further into the galaxy voids.    
Taken from \citet{Zehavi11}.
}
\end{figure}

\begin{figure}
\begin{center}
\includegraphics[scale=0.6]{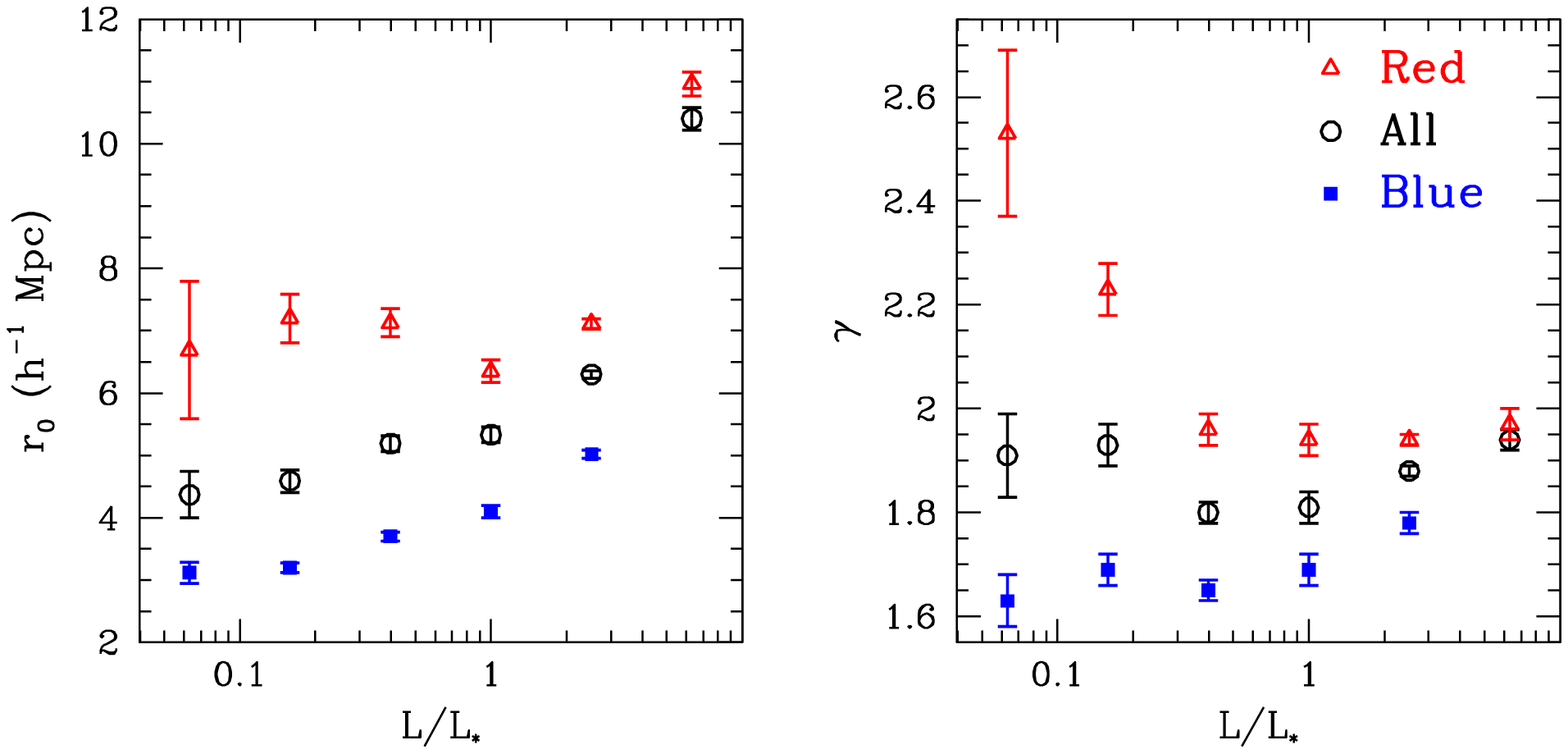}
\end{center}
\caption{The clustering scale length, $r_0$ (left), and slope, $\gamma$ (right), for
all, red, and blue galaxies in SDSS as a function of luminosity.  While all galaxies
are more clustered at brighter luminosities, and red galaxies
are more clustered than blue galaxies at all luminosities, below
$L^*$ the red galaxy clustering length increases at fainter luminosities. The 
clustering slope for faint red galaxies is also much steeper than at other luminosities.
Taken from \citet{Zehavi11}.
}
\end{figure}

Galaxy clustering also depends on other galaxy properties such as
stellar mass, concentration index, and the strength of the 4000\AA
\ break ($D_{4000}$), in that galaxies that have larger stellar mass,
more centrally concentrated light profiles, and/or larger $D_{4000}$
measurements (indicating older stellar populations) are more clustered
\citep{Li06}. This is not surprising given the observed trends with
luminosity and color and the known dependencies of other galaxy
properties with luminosity and color.  Clearly the galaxy bias is a
complicated function of various galaxy properties.

\subsection{Redshift Space Distortions}

\begin{figure}
\begin{center}
\includegraphics[scale=0.6,angle=270,trim=0cm 0cm 0cm 8cm, clip=true]{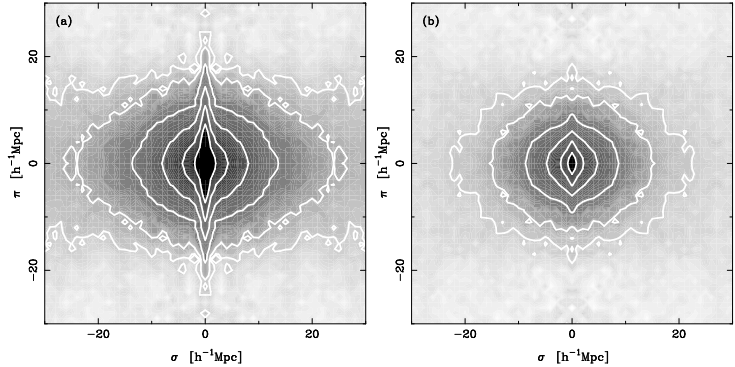}
\end{center}
\caption{Two-dimensional redshift space correlation function \xisp \ 
 (as in Fig.~6 here $\sigma$ is
used instead of $r_p$) for quiescent, absorption line galaxies on the left 
and star forming, emission line galaxies on the right.  
The redshift space distortions are different for the different galaxy populations,
with quiescent and/or red galaxies showing more pronounced ``Fingers of God''.  Both galaxy types exhibit coherent infall on large scales.  
Contours show lines 
of constant $\xi$ at $\xi=$10, 5, 2, 1, 0.5, 0.2, 0.1.  Taken from \citet{Madgwick03}.
}
\end{figure}

The fact that red galaxies are more clustered than blue galaxies is
related to the morphology-density relation \citep{Dressler80}, which
results from the fact that galaxies with elliptical morphologies are
more likely to be found in regions of space with a higher local
surface density of galaxies.  The redshift space distortions seen for
red and blue galaxies also show this.

As discussed in Section 4, redshift space distortions arise from two
different phenomena: virialized motions of galaxies within collapsed
overdensities such as groups and clusters, and the coherent streaming
motion of galaxies onto larger structures that are still collapsing.
The former is seen on relatively small scales ($r_p\lesssim 1$ \mpch)
while the latter is detected on larger scales ($r_P\gtrsim 1$ \mpch).
While the presence of redshift space distortions complicates the
measurement of the real space \xir, these distortions can be used to
uncover information about the thermal motions of galaxies in groups
and clusters as well as the amplitude of the mass density of the
Universe, $\Omega_{\rm matter}$.

Fig.~9 shows \xisp \ for quiescent and star forming galaxies in 2dF.
The quiescent galaxies on the left show larger ``Fingers of God'' than
the star forming galaxies on the right, reflecting the fact that red,
quiescent galaxies have larger motions relative to each others.  This
naturally arises if red, quiescent galaxies reside in more massive,
virialized overdensities with larger random peculiar velocities than
star forming, optically blue galaxies.  The large scale coherent
infall of galaxies is seen both for blue and red galaxies, though it
is often easier to see for blue galaxies, due to their smaller
``Fingers of God''.

These small scale redshift space distortions can be quantified using 
the \sig \ statistic, known as the pairwise velocity dispersion 
\citep{Davis78,Fisher94}.  This is
measured by modeling \xisp \ in real space, which is then 
convolved with a distribution of random pairwise motions, $f(v)$, such that
\begin{equation}
\xi(r_p,\pi) = \int_{-\infty}^{\infty}\xi'(r_p,\pi-v/H_0) f(v) dv,
\end{equation}
where the random motions are often taken to have an exponential form, 
which has been found to fit observed data well:
\begin{equation}
f(v) = \frac{1}{\sigma_{12}\sqrt{2}} \ {\rm exp}\left(- \frac{\sqrt{2}|v|}{\sigma_{12}}\right).
\end{equation}

In the 2dFGRS \citet{Madgwick03} find that \sig$=416 \pm76$ \kms \ for
star forming galaxies and \sig$=612 \pm92$ \kms \ for quiescent
galaxies, measured on scales of 8-20 \mpch.  Using SDSS data
\citet{Zehavi02} find that \sig is $\sim300-450$ \kms \ for blue, star
forming galaxies and $\sim650-750$ \kms \ for red, quiescent galaxies.
It has been shown, however, that this statistic can be sensitive to
large, rare overdensities, such that samples covering large volumes
are needed to measures it robustly.

\citet{Madgwick03} further measure the large scale anisotropies seen
in \xisp \ for galaxies split by spectral type and find that
$\beta=0.49 \pm0.13$ for star forming galaxies and $\beta=0.48
\pm0.14$ for quiescent galaxies.  This implies a similar bias for both
galaxy types on large scales, though they find that on smaller scales
integrated up to 8 \mpch, the relative bias of quiescent to star
forming galaxies is $b_{rel}=1.45 \pm0.14$.

\section{The Evolution of Galaxy Clustering}

The observed clustering of galaxies is expected to evolve with time,
as structure continues to grow due to gravity.  The exact evolution
depends on cosmological parameters such as $\Lambda$ and $\Omega_{\rm
  matter}$.  Larger values of $\Lambda$, for example, lead to larger
voids and higher density contrasts between overdense and underdense
regions.  By measuring the clustering of galaxies at higher redshift,
one can break degeneracies that exist between the galaxy bias and
cosmological parameters that are constrained by low redshift
clustering measurements.  It is therefore useful to determine the
clustering of galaxies as a function of cosmic epoch, not only to
further constrain cosmological parameters but also galaxy evolution.

One might expect the galaxy clustering amplitude \rr \ to increase
over time, as overdense regions become more overdense as galaxies move
towards groups and clusters due to gravity.  However, the exact
evolution of the clustering of galaxies depends not only on gravity,
but also on the expansion history of the Universe and therefore
cosmological parameters such as $\Lambda$.  Additionally, over time
new galaxies form while existing galaxies grow in both mass and
luminosity.  Therefore, the expected changes of galaxy clustering as a
function of redshift depend both on relatively well-known cosmological
parameters and more unknown galaxy formation and evolution physics
which likely depends on gas accretion, star formation, and feedback
processes, as well as mergers.

For a given cosmological model, one can predict how the clustering of
dark matter should evolve with time using cosmological N-body
simulations.  For a \lcdm Universe, \rr \ for dark matter particles is
expected to increase from $\sim$0.8 \mpch \ at $z=3$ to $\sim$5 \mpch
\ at $z=0$ \citep{Weinberg04}.  However, according to hierarchical
structure formation theories, at high redshifts the first galaxies to
form will be the first structures to collapse, which will be biased
tracers of the mass.  The galaxy bias is expected to be a strong
function of redshift, initially $>1$ at high redshift and approaching
unity over time.  Therefore, \rr \ for galaxies may be a much weaker
function of time than it is for dark matter, as the same galaxies are
not observed as a function of redshift, and over time new galaxies
form in less biased locations in the Universe.

The projected angular and three dimensional correlation functions of
galaxies have been observed to $z\sim5$.  Star-forming Lyman break
galaxies at $z\sim3-5$ are found to have \rr \ $\sim4-6$ \mpch, with a
bias relative to dark matter of $\sim3-4$ \citep{Ouchi04,
  Adelberger05}.  Brighter Lyman break galaxies are found to cluster
more strongly than fainter Lyman break galaxies.  The correlation
length, \rr, is found to be roughly constant between $z=5$ and $z=3$,
implying that the bias is increasing at earlier cosmic epoch.
Spectroscopic galaxy surveys at $z>2$ are currently limited to samples
of at most a few thousand galaxies, so most clustering measurements
are angular at these epochs.  In one such study by \citet{Wake11},
photometric redshifts of tens of thousands of galaxies at $1 < z < 2$
are used to measure the angular clustering as a function of stellar
mass.  They find a strong dependence of clustering amplitude on
stellar mass in each of three redshift intervals in this range.

At $z\sim1$ larger spectroscopic galaxy samples exist, and three
dimensional two-point clustering analyses have been performed as a
function of luminosity, color, stellar mass, and spectral type.  The
same general clustering trends with galaxy property that are observed
at $z\sim0$ are also seen at $z\sim1$, in that galaxies that are
brighter, redder, early spectral type, and/or more massive are more
clustered.  The clustering scale length of red galaxies is found to be
$\sim5-6$ \mpch \ while for blue galaxies it is $\sim3.5-4.5$ \mpch,
depending on luminosity \citep{Meneux06, Coil08}.  At a given
luminosity the observed correlation length is only $\sim$15\% smaller
at $z=1$ than $z=0$, indicating that unlike for dark matter the galaxy
\rr \ is roughly constant over time. These results are consistent with
predictions from \lcdm simulations.

The measured values of \rr \ at $z\sim1$ imply that are more biased at
$z=1$ than at $z=0$.  Within the DEEP2 sample, the bias measured on
scales of $\sim1-10$ \mpch \ varies from $\sim1.25-1.55$, with
brighter galaxy samples being more biased tracers of the dark matter
\citep{Coil06}.  These results are consistent with the idea that
galaxies formed early on in the most overdense regions of space, which
are biased.

As in the nearby Universe, the clustering amplitude is a stronger
function of color than of luminosity at $z\sim1$.  Additionally, the
color-density relation is found to already be in place at $z=1$, in
that the relative bias of red to blue galaxies is as high at $z=1$ as
at $z=0.1$ \citep{Coil08}.  This implies that the color-density
relation is not due to cluster-specific physics, as most galaxies at
$z=1$ in field spectroscopic surveys are not in clusters.  Therefore
it must be physical processes at play in galaxy groups that initially
set the color and morphology-density relations.  Red galaxies show
larger ``Fingers of God'' in \xisp \ measurements than blue galaxies
do, again showing that red galaxies at $z=1$ lie preferentially in
virialized, more massive overdensities compared to blue galaxies.
Both red and blue galaxies show coherent infall on large scales.

\section{Halo Model Interpretation of \xir}

The current paradigm of galaxy formation posits that galaxies form in
the center of larger dark matter halos, collapsed overdensities in the
dark matter distribution with $\rho / \bar{\rho} \sim200$, inside of
which all mass is gravitationally bound.  
The clustering of galaxies can then be understood as a combination of 
the clustering of dark matter halos, which depends on cosmological 
parameters, and how galaxies populate dark matter halos, which depends 
on galaxy formation and evolution physics.  
For a given cosmological
model the properties of dark matter halos, including their evolution
with time, can be studied in detail using N-body simulations.  The
masses and spatial distribution of dark matter halos should depend
only on the properties of dark matter, not baryonic matter, and the
expansion history of the Universe; therefore the clustering of dark
matter halos should be insensitive to baryon physics.  However, the
efficiency of galaxy formation is very dependent on the complicated
baryonic physics of, for example, star formation, gas cooling, and
feedback processes.  The halo model allows the relatively simple
cosmological dependence of galaxy clustering to be cleanly separated
from the more complex baryonic astrophysics, and it shows how
clustering measurements for a range of galaxy types can be used to
constrain galaxy evolution physics.

\subsection{Estimating the Mean Halo Mass from the Bias}

One can use the observed large scale clustering amplitude of different
observed galaxy populations to identify the typical mass of their
parent dark matter halos, in order to place these galaxies in a
cosmological context.  The large scale clustering amplitude of dark
matter halos as a function of halo mass is well determined in N-body
simulations, and analytic fitting formula are provided by e.g.,
\citet{Mo96} and \citet{Sheth01}.  Analytic models can then predict
the clustering of both dark matter particles and galaxies as a function
of scale, by using the clustering of dark matter halos and the radial
density profile of dark matter and galaxies within those halos
\citep{Ma00, Peacock00, Seljak00}.  In this scheme, on large, linear
scales where $\delta<1$ ($\rho / \bar{\rho} \sim1$), the clustering of
a given galaxy population can be used to determine the mean mass of
the dark matter halos hosting those galaxies, for a given cosmological
model.  
To achieve this, the large-scale bias is estimated by comparing the
observed galaxy clustering amplitude with that of dark matter in an N-body  
simulation, and then galaxies are assumed to reside in halos of a
given mass that have the same bias in simulations.  

Simulations show that higher mass halos  cluster more 
strongly than lower mass halos \citep{Sheth99}.  This then leads 
to an interpretation
of galaxy clustering as a function of luminosity in which luminous
galaxies reside in more massive dark matter halos than less luminous
galaxies.  Similarly, red galaxies typically reside in more massive
halos than blue galaxies of the same luminosity; this is 
observationally verified by the
larger ``Fingers of God'' observed for red galaxies.  Combining the
large scale bias with the observed galaxy number density further
allows one to constrain the fraction of halos that host a given galaxy
type, by comparing the galaxy space density to the parent dark matter
halo space density. This constrains the duty cycle or fraction of 
halos hosting galaxies of a given population.

\subsection{Halo Occupation Distribution Modeling}

While such estimates of the mean host halo mass and duty cycle are 
fairly straightforward to carry out, a greater understanding of 
the relation between galaxy light and dark matter mass 
is gleaned by performing halo occupation 
distribution modeling.

The general halo-based model discussed above, in which the clustering of galaxies
reflects the clustering of halos, was further developed by
\citet{Peacock00} to include the efficiency of galaxy formation, or
how galaxies populate halos.  The proposed model depends on both the
halo occupation number, equal to the number of galaxies in a halo of a
given mass, for a galaxy sample brighter than some limit, and the
location of the galaxies within these halos.  In the \citet{Peacock00}
model it is assumed that one galaxy is at the center of the halo (the
``central'' galaxy), and the rest of the galaxies in the same halo are
``satellite'' galaxies that trace the dark matter radial mass
distribution, which follows an NFW profile \citep{Navarro97}.  The
latter assumption results in a general power law shape for the galaxy
correlation function.

A similar idea was proposed by \citet{Benson00}, who used a
semi-analytic model in conjunction with a cosmological N-body
simulation to show that the observed galaxy \xir \ could be reproduced
with a \lcdm \ simulation (though not with a $\tau$ CDM simulation
with $\Omega_{\rm matter}=1$).  They also employ a method for locating
galaxies inside dark matter halos such that one galaxy resides at the
center of all halos above a given mass threshold, while additional
galaxies are assigned the location of a random dark matter particle
within the same halo, such that galaxies have the same NFW radial profile
within halos as the dark matter particles (see Fig.~10, left panel).

In these models, the clustering of galaxies on scales larger
than a typical halo ($\sim1-2$ \mpch) results from pairs of galaxies
in separate halos, called the ``two-halo term'', while the clustering
on smaller scales ($\lesssim1$ \mpch) is due to pairs of galaxies within
the same parent halo, called the ``one-halo term''. 
When the pairs from these two terms are added together, the 
 resulting galaxy correlation function should roughly
follow a power law.

\citet{Benson00} show that on large scales there is a simple relation in the
bias between galaxies and dark matter halos, while on small scales the
correlation function depends on the number of galaxies in a halo and
the finite size of halos.  When the clustering signal from these two
scales 
(corresponding to the ``two-halo'' and ``one-halo'' terms)
 is combined, a power law results for the galaxy \xir \ (right panel,
Fig.~10).  Galaxies are found to be anti-biased relative to dark
matter (i.e., less clustered than the dark matter) 
on scales smaller than the typical halo, though the bias is
close to unity on larger scales.  The clustering of galaxies that
results from this semi-analytic model is also found to match the observed 
clustering of galaxies in the APM survey, above a given luminosity threshold
\citep{Baugh96}.

\begin{figure}
\begin{center}
\includegraphics[width=0.44\linewidth]{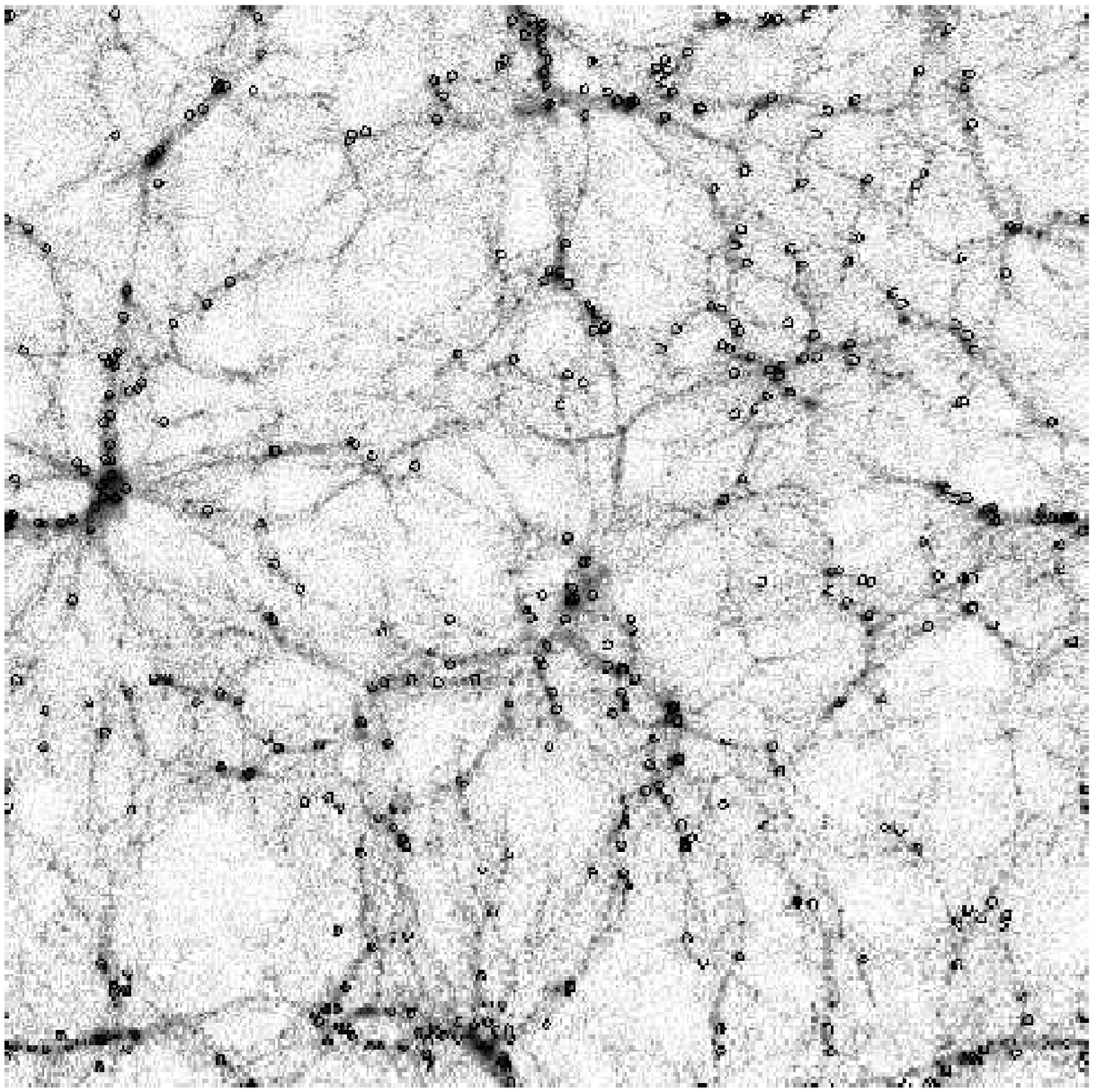}
\includegraphics[width=0.47\linewidth]{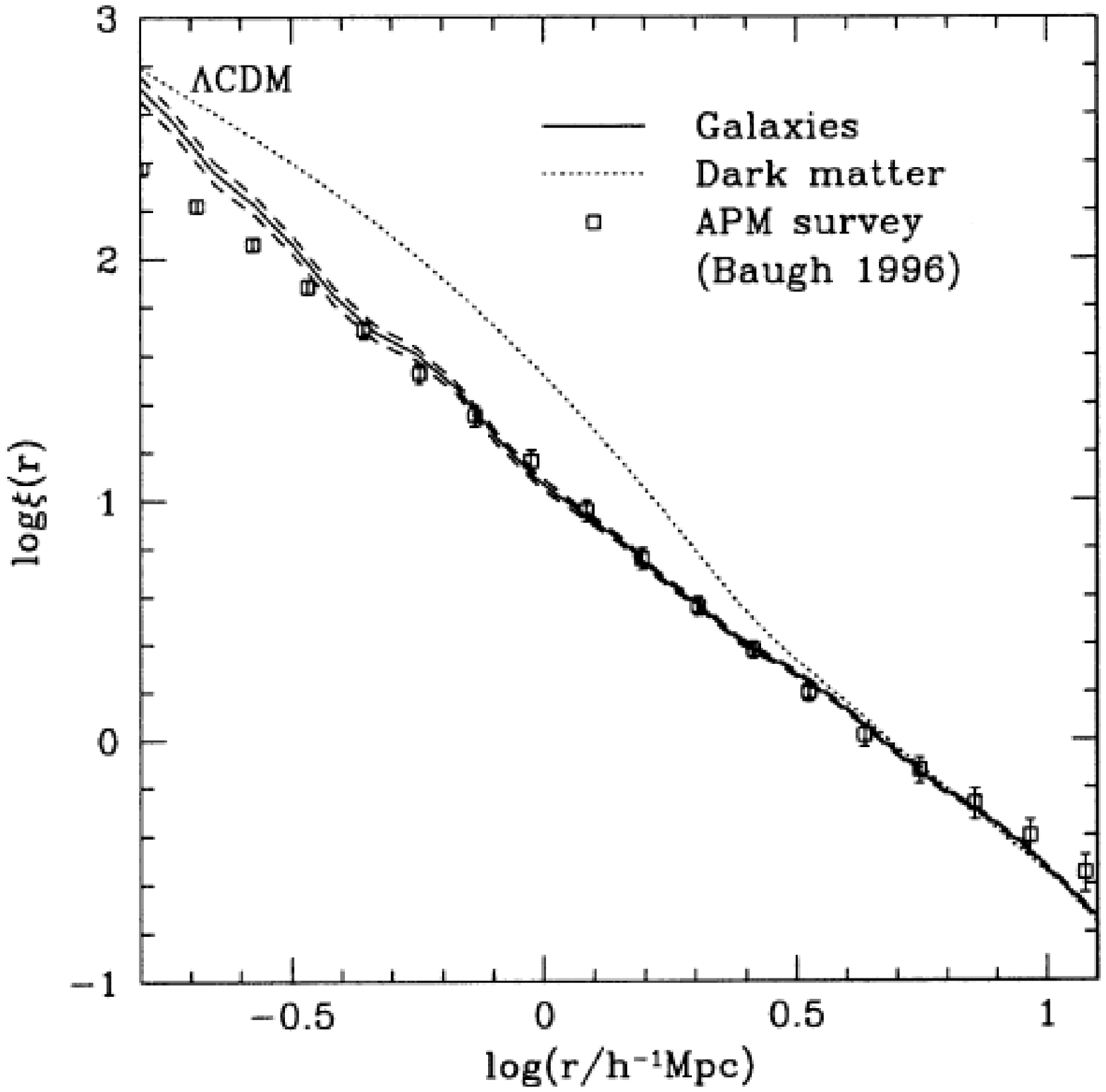}
\end{center}
\caption{Left: The large scale structure seen in a \lcdm \ N-body dark matter 
only 
simulation of size 141 $\times$ 141 $\times$ 8 $h^{-3} Mpc^3$.  The grey scale 
indicates the density of dark matter, while the locations of galaxies are
shown with open circles.  Galaxies are added to the simulation output using 
a semi-analytic model which assumes that dark matter halos above a given mass 
threshold have at least one ``central'' galaxy located at the center of
the halo.  Higher mass halos contain additional ``satellite'' galaxies, which
are assigned the location of a random dark matter particle in the halo. 
Taken from \citet{Benson00}.
Right: The two-point correlation function of dark matter particles 
(dotted line) 
and galaxies (solid line with dashed line showing Poisson error bars) 
in the simulation of \citet{Benson00}, compared with the observed clustering of 
galaxies in the APM survey (open squares) \citep{Baugh96}. 
}
\end{figure}

By defining the halo occupation distribution (HOD) as the probability
that a halo of a given mass contains N galaxies, $P(N|M)$,
\citet{Berlind02} quantify how the observed galaxy \xir \ depends on
different HOD model parameters.  Using N-body simulations, they
identify dark matter halos and place galaxies into the simulation
using a simple HOD model with two parameters: a minimum mass at which
a halo hosts, on average, one central galaxy ($M_{min}$) at the center
of the halo, and the slope
($\alpha$) of the $P(N|M)$ function for satellite galaxies.  The
latter determines how many satellite galaxies there are as a
function of halo mass.  They further assume that the satellite 
galaxies follow an
NFW profile, as the dark matter does, though the concentration of the
radial profile can be changed.  They show that the ``two-halo term''
is simply the halo center correlation function weighted by a large
scale bias factor, while the ``one-halo term'' is sensitive to both
$\alpha$ and the concentration of the galaxy profile within
halos. Obtaining a power law \xir \ therefore strongly constrains the
HOD model parameters.

\citet{Kravtsov04} propose that the locations of satellite galaxies within dark 
matter halos should correspond to locations of subhalos, distinct gravitationally 
bound regions within the larger dark matter halos, instead of tracing
random dark matter particles.  Using cosmological N-body 
simulations, they show that at $z>1$ \xir \ for galaxies should deviate 
strongly from 
a power law on small scales, due to a rise in the ``one-halo term''.   
In this model, the clustering of galaxies can be understood as the 
clustering of dark matter parent halos and subhalos, and the power law 
shape that is 
observed at $z\sim0$ is a coincidence of the one- and two-halo terms having 
similar amplitudes and slopes at the typical scale of halos.
They find that the formation and evolution of halos and subhalos through 
merging and dynamical processes are the main physical drivers of large scale
structure.

With the unprecedentedly large galaxy sample with spectroscopic redshifts 
that is provided by SDSS, departures from a power law \xir \ were detected by 
\cite{Zehavi04}, using a volume-limited subsample of 22,000 galaxies from a 
parent sample of 118,000 galaxies.  The deviations from a power law are small 
enough at $z\sim0$ that 
a large sample covering a sufficiently large cosmological volume is 
required to overcome the errors due to cosmic variance to detect 
these small deviations.
It is found that there is a change in 
the slope of \xir \ on scales of $\sim$1-2 \mpch; this 
corresponds to the scale at which the one and two halo term are equal (see
Fig.~11). 
\cite{Zehavi04} find that \xir \ measured from the SDSS data is better fit 
by an HOD model, which includes small deviations from a power law, 
than by a pure power law.  The HOD model that is fit has three parameters:
the minimum mass to host a single central galaxy ($M_{\rm min}$), 
the minimum mass to host a single satellite galaxy ($M_{\rm 1}$), and the 
slope of  $P(N|M)$ ($\alpha$), which determines the average number of 
satellite galaxies as a function of host halo mass.  In this model, dark matter
halos with $M_{\rm min} < M  < M_{\rm 1}$ host a single galaxy, while above 
$M_{\rm 1}$ they host, on average, $(M/M_{\rm 1})^{\alpha}$ galaxies.
Using \wprp, one can fit for $M_{\rm 1}$ and $\alpha$, while the observed 
space density of galaxies is used  to derive $M_{\rm min}$. For a galaxy
sample with $M_r<-21$, the best-fit HOD parameters are
 $M_{\rm min}=6.1 \times 10^{12}$ \hmsun, 
$M_{\rm 1}=4.7 \times 10^{13}$ \hmsun, and $\alpha=0.89$.

\begin{figure}
\begin{center}
  \includegraphics[width=0.45\linewidth]{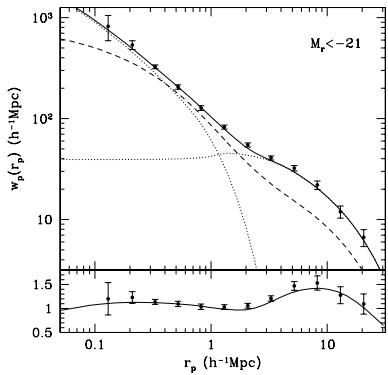}
\end{center}
\caption{The projected correlation function, \wprp, for SDSS galaxies with 
$M_r<-21$ is shown as data points with error bars.  The best-fit HOD model 
is shown as a solid line, with the contributions from the one and two halo 
terms shown with dotted lines.  The projected correlation function of dark
matter at this redshift is shown with a dashed line. 
The bottom panel shows deviations in \wprp \ for the data and the HOD
model from the best-fit power law.
Taken from \citet{Zehavi04}.
}
\end{figure}

\subsection{Interpreting the Luminosity and Color Dependence of Galaxy Clustering}

In general, these HOD parameters reflect the efficiency of galaxy formation 
and evolution and can be a function of galaxy properties such as luminosity, 
color, stellar mass, and morphology.  \cite{Zehavi11} present
HOD fits to SDSS samples as a function of luminosity and color and find that 
$\alpha$ is generally $\sim$1.0-1.1, though it is a bit higher for the 
brightest galaxies ($\sim$1.3 for $M_r<-22.0$).
There is a strong trend between luminosity and halo mass; 
$M_{\rm min}$ varies as a function of luminosity from 
$\sim 10^{11}$ \hmsun \ for $M_r<-18$ to $\sim10^{14}$ \hmsun \ for $M_r<-22$.
$M_{\rm 1}$ is generally $\sim$17 times higher than the value of $M_{\rm min}$ 
for all luminosity threshold samples (see Fig.~12).  
This implies that a halo with two 
galaxies above a given luminosity is $\sim$17 times more
massive than a halo hosting one galaxy above the same luminosity limit.  
Further, the fraction of galaxies that are 
satellites decreases at higher luminosities, from $\sim$33\% at $M_r<-18$ to 
4\% at $M_r<-22$.  The right panel of Fig.~12 shows the mass-to-light ratio
of the virial halo mass to the central galaxy $r$-band luminosity as a function
of halo mass.  This figure shows that halos of mass $\sim4 \times 10^{11}$ 
\hmsun  \ are maximally efficient at galaxy formation, at converting baryons 
into light.

\begin{figure}
\begin{center}
  \includegraphics[width=0.9\linewidth]{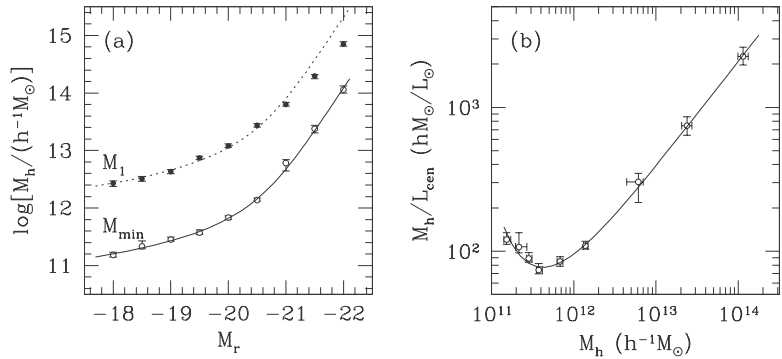}
\end{center}
\caption{Left: The characteristic mass scale of dark matter halos hosting 
galaxies as a function of the luminosity threshold of the galaxy sample. Both
the minimum halo mass to host a single galaxy is shown ($M_{\rm min}$) as well
as the minimum mass to host additional satellite galaxies ($M_{\rm 1}$). 
 A strong 
relationship clearly exists between halo mass and galaxy luminosity.
Right:  The ratio of the halo mass to the median central galaxy luminosity 
as a function of halo mass.  
Taken from \citet{Zehavi11}.
}
\end{figure}

In terms of the color dependence of galaxy clustering, 
 the trend at fainter luminosities of red galaxies 
being strongly clustered (with a higher correlation slope, $\gamma$, see 
Fig.~8) is due to faint red galaxies being satellite galaxies in relatively 
massive halos that host bright red central galaxies \citep{Berlind05}.
HOD modeling therefore provides a clear explanation for the increased 
clustering observed for faint red galaxies.
For a given luminosity range $(-20 < M_r < -19)$ \cite{Zehavi11}   
fit a simplified HOD model with one parameter only to find that the fraction 
of galaxies that are satellites is much higher for red than for blue galaxies, 
with $\sim$25\% of blue galaxies being satellites and $\sim$60\% of red 
galaxies being satellites.  
They find that blue galaxies reside in halos with a median mass of 
$10^{11.7}$ \hmsun, while red galaxies reside in higher mass halos with a median
mass of  $10^{12.2}$ \hmsun. 
However, at a given luminosity, there is not a strong 
trend between color and halo mass (though there is a strong trend between
luminosity and halo mass). Instead, the differences in \wprp \ reflect a 
trend between color and satellite fraction; the increased satellite fraction, 
in particular, drives the slope of \xir \ to be steeper for red galaxies 
compared to blue galaxies.  And while the HOD slope $\alpha$, does not 
change much with increasing luminosity, it does with color, due to the 
dependence of the satellite fraction on color.  
Having a higher satellite fraction also 
places more galaxies in high mass halos (as those host the groups and clusters
 that contain the satellite galaxies), which increases the large scale bias 
and boosts the one halo term relative to the two halo term.
The HOD model facilitates interpretion of the observed luminosity and color 
dependence of galaxy clustering and provides strong, crucial constraints 
on models of how galaxies form and evolve within their parent dark matter halos.

\subsection{Interpreting the Evolution of Galaxy Clustering}

As mentioned in Section 7 above, the galaxies that are observed for clustering 
measurements
at different redshifts are not necessarily the same populations across cosmic
time.  A significant hurdle in understanding galaxy evolution is knowing how 
to connect different observed populations at different redshifts.  Galaxy 
clustering measurements can be combined with theoretical models 
to trace observed populations with 
redshift, in that for a given cosmology one can model how the clustering of
a given population should evolve with time.  

The observed evolution of the luminosity-dependence of galaxy clustering 
can be fit surprisingly well using a simple non-parametric, non-HOD, 
model that relates
the galaxy luminosity function to the halo mass function.  \citet{Conroy06}
show that directly matching galaxies as a function of luminosity to host halos
and subhalos 
as a function of mass leads to a model for the luminosity-dependent clustering 
that matches observation from $z\sim0$ to $z\sim3$.  In this model, the only 
inputs are the observed galaxy luminosity function at each epoch of interest 
and the dark matter halo (and subhalo) 
mass function from N-body simulations.  Galaxies are 
then ranked by luminosity and halos by mass and matched one-to-one, such 
that lower luminosity galaxies are associated with halos of lower mass, and
galaxies above a given luminosity threshold are assigned to halos above a given
mass threshold with the same abundance or number density.  This ``abundance 
matching'' method uses as a proxy for halo mass the maximum circular velocity 
($V_{\rm max}$) of the halo; for subhalos they find that it is necessary to 
use the value of $V_{\rm max}$ when the subhalo is first accreted into a larger 
halo, to avoid the effects of tidal stripping.  With this simple model the 
clustering amplitude and shape as a function of luminosity are matched for
SDSS galaxies at $z\sim0$, DEEP2 galaxies at $z\sim1$, and Lyman break 
galaxies at $z\sim3$.  In particular, the clustering amplitude in both 
the one and two halo regimes is well fit, including the deviations from a 
power law that seen at $z>1$ \citep{Ouchi05,Coil06}.  These results imply 
a tight correlation between galaxy luminosity and halo mass from $z\sim0$ 
to $z\sim3$.

While abundance-matching techniques provide a simple, zero parameter model
for how galaxies populate halos, a richer understanding of the physical 
properties involved may be gained by performing HOD modeling.  
\citet{Zheng07} use HOD modeling to fit the observed 
luminosity-dependent galaxy clustering at $z\sim0$ 
measured in SDSS with that measured at $z\sim1$ in DEEP2 to confirm that at 
both epochs there is a tight relationship between the central galaxy 
luminosity and host halo mass.  At $z\sim1$ the satellite fraction drops 
for higher luminosities, as at $z\sim0$, but at a given luminosity the 
satellite fraction is higher at $z\sim0$ than at $z\sim1$.  They also find
that at a given central luminosity, halos are $\sim$1.6 times more massive 
at $z\sim0$ than $z\sim1$, and at a given halo mass galaxies are $\sim$1.4 
times more luminous at $z\sim1$ than $z\sim0$.

\citet{Zheng07} further combine these HOD results with theoretical predictions of 
the growth of dark matter halos from simulations to link $z\sim1$ central 
galaxies to their descendants at $z\sim0$ and find that the growth of both
halo mass and stellar mass as a function of redshift 
depends on halo mass.  Lower mass halos grow 
earlier, which is reflected in the fact that more of their $z\sim0$ mass 
is already assembled by $z\sim1$.  A typical $z\sim0$ halo with mass $3 \times 
10^{11}$ \hmsun \ has about 70\% of its final mass in place by $z\sim1$, 
while a $z\sim0$ halo with mass $10^{13}$ \hmsun \ has $\sim$50\% of its 
final mass in place at $z\sim1$.  In terms of stellar mass, however, 
in a $z\sim0$ halo of mass $5 \times 10^{11}$ \hmsun \ a central galaxy has 
$\sim$20\% of its stellar mass in place at $z\sim1$, while the fraction 
rises to $\sim$33\% above a halo mass of $2 \times 10^{12}$ \hmsun. They
further find that the mass scale of the maximum star formation efficiency 
for central galaxies shifts to lower halo mass with time, with a peak of 
$\sim10^{12}$ \hmsun \ at $z\sim1$ and $\sim6 \times 10^{11}$ \hmsun at $z\sim0$.

At $1<z<2$, \citet{Wake11} use precise photometric redshifts from the 
NEWFIRM survey to measure the 
relationship between stellar mass and dark matter halo mass using HOD models.  
At these higher redshifts \rr \ varies from $\sim$6 to $\sim$11 \mpch \ for
 stellar 
masses $\sim10^{10}$ \msun \ to $10^{11}$ \msun.  The galaxy bias is a 
function of both redshift and stellar mass and is $\sim$2.5 at $z\sim1$ 
and increases to $\sim$3.5 at $z\sim2$. They find that the typical halo mass 
of both central and satellite galaxies increases with stellar mass, while 
the satellite fraction drops at higher stellar mass, qualitatively similar 
to what is found at lower redshift.  They do not find evolution in the 
relationship between stellar mass and halo mass between $z\sim2$ and $z\sim1$, 
but do find evolution compared to $z\sim0$.  
They also find that the peak of star formation efficiency shifts to 
lower halo mass with time.

Simulations can also be used to connect different observed galaxy 
populations at different redshifts.  An example of the power of this method 
is shown by \cite{Conroy08}, who compare the clustering and space density 
of star forming galaxies at $z\sim2$ 
with that of star forming and quiescent galaxies at $z=1$ and $z=0$ to infer 
both
the typical descendants of the $z\sim2$ star forming galaxies  and constrain 
the fraction that have merged with other galaxies by $z=0$. 
They use halos and subhalos identified in a \lcdm \ N-body simulation to 
determine which halos at $z\sim2$ likely host star forming galaxies, and 
then use the merger histories in the simulation to track these same halos
to lower redshift.  By 
comparing these results to observed clustering of star forming galaxies at 
$z\sim1$ and $z\sim0$ they can identify the galaxy populations at these 
epochs that are consistent with being descendants of the $z\sim2$ galaxies. 
They find that while the lower redshift descendent halos have clustering 
strengths similar to red galaxies at both $z\sim1$ and $z\sim0$, the $z\sim2$ 
star forming 
galaxies can not all evolve into red galaxies by lower redshift, as their 
space density is too high.  There are many more lower redshift descendents
 than there are red galaxies, even after taking into account mergers.  They 
conclude that most $z\sim2$ star forming galaxies evolve into typical
$L^*$ galaxies today, 
while a non-negligible fraction become satellite galaxies in larger galaxy 
groups and clusters.

In summary, N-body simulations and HOD modeling 
can be used to interpret the observed evolution of galaxy clustering 
and further constrain both cosmological parameters and theoretical models of 
galaxy evolution beyond what can be gleaned from $z\sim0$ observations alone.  
They 
also establish links between distinct observed galaxy populations at different 
redshifts, allowing one to create a coherent picture of how galaxies 
evolve over cosmic time.

\section{Voids and Filaments}

Redshift surveys unveil a rich structure of galaxies, as seen in Fig.~3.  In
addition to measuring the two-point correlation function to quantify the 
clustering amplitude as a function of galaxy properties, one can also study
higher-order clustering measurements as well as properties of voids and 
filaments.

\subsection{Higher-order Clustering Measurements}

Higher-order clustering statistics reflect both the growth of initial density 
fluctuations as well as the details of galaxy biasing \citep{Bernardeau02}, 
such that measurements of higher-order clustering can test the paradigm of 
structure formation through gravitational instability as well as constrain 
the galaxy bias.  
In the linear regime there is a degeneracy between the amplitude of 
fluctuations in the dark matter density field and the galaxy bias, in that a highly 
clustered galaxy population may be biased and trace only the most overdense regions 
of the dark matter, or the dark matter itself may be highly clustered.  However, 
this degeneracy can be broken in the non-linear regime on small scales.  Over 
time, the density field becomes skewed towards high density as 
$\delta$ becomes greater than unity in overdense regions 
(where $\delta \equiv (\rho / \bar{\rho}) -1$ )
but can not become negative in 
underdense regions.  Skewness in the galaxy density distribution 
can also arise from galaxy bias, if galaxies preferentially form in the 
highest density peaks.  One can therefore use the shapes of the galaxy 
overdensities, through measurements of the three-point correlation function, 
to test gravitational collapse versus galaxy bias.

To study higher-order clustering one needs large samples that cover
enormous volumes; all studies to date have focused on low redshift galaxies.
\citet{Verde02} use 2dFGRS to measure the Fourier transform of the three-point
correlation function, called the bispectrum, to constrain the galaxy bias 
without resorting to comparisons with N-body simulations in order to measure 
the clustering of dark matter.  
\citet{Fry93} present the galaxy bias in terms of a Taylor expansion of the 
density contrast, where the first order term is the linear term, while the 
second order term is the non-linear or quadratic term.  
Measured on scales of 5 -- 30 \mpch, \citet{Verde02} 
find that the linear galaxy bias is consistent with unity ($b_1 = 1.04 \pm0.11$), 
while the non-linear quadratic bias is consistent with zero 
($b_2 = -0.05 \pm0.08$).  When combined with the redshift space distortions 
measured in the two-dimensional two-point correlation function (\xisp),
they measure $\Omega_{\rm matter} = 0.27 \pm0.06$ at $z=0.17$.  This constraint 
on the matter density of the Universe is derived entirely from large scale
structure data alone.

\citet{Gaztanaga05} measure the three-point correlation function in 2dFGRS 
for triangles of galaxy configurations with different shapes.  
Their results are consistent with \lcdm expectations regarding gravitational 
instability of initial Gaussian fluctuations.  Furthermore, 
they find that while the 
linear bias is consistent with unity ($b_1=0.93 +0.10/-0.08$), the 
quadratic bias is non-zero ($b_2/b_1 = -0.34 +0.11/-0.08$).  This implies that 
there is a non-gravitational contribution to the three-point function, resulting from
galaxy formation physics.  These
results differ from those of \citet{Verde02}, which may be due to the inclusion 
by \citet{Gaztanaga05} of the covariance between measurements on different scales.  
\citet{Gaztanaga05} combine their results with the measured two-point 
correlation function to derive $\sigma_8 = 0.88 +0.12/-0.10$.

If the density field follows a Gaussian distribution, 
the higher-order clustering terms can be expressed solely in terms of the 
lower order clustering terms.  This ``hierarchical scaling'' holds 
for the evolution of an initially Gaussian distribution of fluctuations
 under gravitational instability.  Therefore departures from hierarchical 
scaling can result either from a non-Gaussian initial density field or 
from galaxy bias.  
Redshift space higher-order clustering measurements in 2dFGRS are performed by
 \citet{Baugh04} and \citet{Croton04b}, who measure up to the six-point
correlation function.  They find that hierarchical scaling is obeyed on 
small scales, though deviations exist on larger scales ($\sim10$ \mpch).  
They show that on large scales the higher-order terms can be significantly 
affected
by massive rare peaks such as superclusters, which populate the tail of the 
overdensity distribution.  \citet{Croton04b} also show that the three-point 
function has a weak luminosity dependence, implying that galaxy bias is not 
entirely linear.  These results are confirmed by \citet{Nichol06} using 
galaxies in the SDSS, who also measure a weak luminosity dependence in the 
three-point function.  They find that on scales $>$10 \mpch \ the three-point
function is greatly affected by the ``Sloan Great Wall'', a massive 
supercluster that is roughly 450 Mpc \citep{Gott05} in length and is 
associated with tens of known Abell clusters.  These results show that even
2dFGRS and SDSS are not large enough samples to be unaffected by the most 
massive, rare structures.

Several studies have examined higher-order correlation functions for galaxies
split by color.  \citet{Gaztanaga05} find a strong dependence of the 
three-point function on color and luminosity on scales $<$6 \mpch. 
\citet{Croton07} measure up to the five-point correlation function in 
2dFGRS for both blue and red galaxies and find that red galaxies are more 
clustered than blue galaxies in all of the N-point functions measured.  
They also find a luminosity-dependence in the hierarchical scaling amplitudes 
for red galaxies but not for blue galaxies.  Taken together, these results 
explain why the full galaxy population shows only a weak correlation with 
luminosity.

\subsection{Voids}

In maps of the large scale structure of galaxies, voids stand out
starkly to the eye.  There appear to be vast regions of space with
few, if any, $L^*$ galaxies.  Voids are among the largest structures
observed in the Universe, spanning typically tens of \mpch.

The statistics of voids -- their sizes, distribution, and
underdensities -- are closely tied to cosmological parameters and the
physical details of structure formation (e.g. \citet{Sheth04}).  
While the two-point 
correlation function provides a full description of clustering for a 
Gaussian distribution, departures from Gaussianity can be tested with 
higher-order correlation statistics and voids.
For example, the abundance
of voids can be used to test the non-Gaussianity of primordial
perturbations, which constrains models of inflation
\citep{Kamionkowski09}.  Additionally, voids provide an extreme
low density environment in which to study galaxy evolution.  As discussed by
\citet{Peebles01}, the lack of galaxies in voids should provide a
stringent test for galaxy formation models.

\begin{figure}
\begin{center}
\includegraphics[scale=0.4,angle=270]{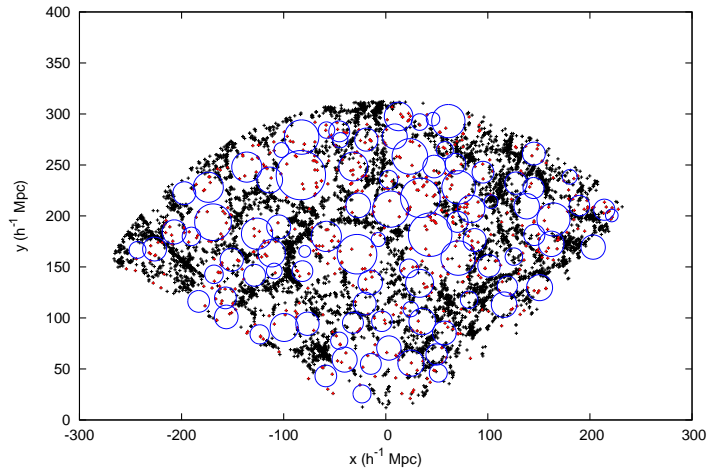}
\end{center}
\caption{Void and wall galaxies in the SDSS.  Shown is a projection of a 
10 \mpch \ slab with wall galaxies plotted as black crosses and 
void galaxies plotted as red crosses.  Blue circles indicate the 
intersection of the maximal sphere of each void with the midplane of
the slab (from Pan et al. 2011).
}
\end{figure}

\subsubsection{Void and Void Galaxy Properties}

The first challenge in measuring the properties of voids and void
galaxies is defining the physical extent of individual voids and
identifying which galaxies are likely to be in voids.  The ``void 
finder'' algorithm of \citet{Elad97}, which is based on the point
distribution of galaxies (i.e., does not perform any smoothing), is
widely used.  This algorithm does not assume that voids are entirely
devoid of galaxies and identifies void galaxies as those with three or
less neighboring galaxies within a sphere defined by the mean and
standard deviation of the distance to the third nearest neighbor for
all galaxies.  All other galaxies are termed ``wall'' galaxies.  An
individual void is then identified as the maximal sphere that contains
only void galaxies (see Fig.~10).  
This algorithm is widely used by both theorists and observers.  

Cosmological simulations of structure formation show that the
distribution and density of galaxy voids are sensitive to the values
of $\Omega_{\rm matter}$ and $\Omega_{\rm \Lambda}$
\citep{Kauffmann99a}.  Using \lcdm N-body dark matter simulations,
\citet{Colberg05} study the properties of voids within the dark
matter distribution and predicts that voids are very underdense
(though not empty) up to a well-defined, sharp edge in the dark matter
density.  They predict that 61\% of the volume of space
should be filled by voids at $z=0$, compared to 28\% at $z=1$ and 9\% at $z=2$.
They also find that the mass function of dark matter halos in voids is
steeper than in denser regions of space.

Using similar \lcdm N-body simulations with a semi-analytic model for galaxy 
evolution, \citet{Benson03} show that voids should contain both 
dark matter and galaxies, and that the dark matter halos in voids 
tend to be low mass and 
therefore contain fewer galaxies than in higher density regions.  
In particular, at density contrasts of $\delta < -0.6$, where 
$\delta \equiv (\rho / \rho_{\rm mean}) - 1$, both dark matter halos and 
galaxies in voids should be anti-biased relative
to dark matter.  However, galaxies are 
predicted to be more underdense than the dark matter halos, assuming 
simple physically-motivate prescriptions for galaxy evolution.
They also predict the statistical size distribution of voids, finding 
that there should be more voids with smaller radii ($<10$\mpch) 
than larger radii.

The advent of the 2dFGRS and SDSS provided the first very large
samples of voids and void galaxies that could be used to robustly
measure their statistical properties.  Applying the "void finder"
algorithm on the 2dFGRS dataset, \citet{Hoyle04} find that the typical
radius of voids is $\sim$15 \mpch.  Voids are extremely underdense,
with an average density of $\delta\rho / \rho$=-0.94, with even lower
densities at the center, where fewer galaxies lie.  The volume of
space filled by voids is $\sim$40\%.  Probing an even larger volume of
space using the SDSS dataset, \citet{Pan11} find a similar typical
void radius and conclude that $\sim$60\% of space is filled by voids,
which have $\delta\rho / \rho$=-0.85 at their edges.  Voids have sharp
density profiles, in that they remain extremely underdense to the void
radius, where the galaxy density rises steeply.
These observational results agree well with the predictions of \lcdm  
simulations discussed above.

Studies of the properties of galaxy in voids allow an understanding of
how galaxy formation and evolution progresses in the lowest density
environments in the Universe, effectively pursuing the other end of
the density spectrum from cluster galaxies. 
Void galaxies are found to be significantly bluer and fainter than
wall galaxies \citep{Rojas04}.  The luminosity function of void
galaxies shows a lack of bright galaxies but no difference in the
measured faint end slope \citep{Croton05, Hoyle05}, indicating that
dwarf galaxies are not likely to be more common in voids. The normalization 
of the luminosity function of wall galaxies is roughly an order of magnitude 
higher than that of void galaxies; therefore
galaxies do exist in voids, just with a much lower space density. Studies of
the optical spectra of void galaxies show that they have high star
formation rates, low 4000\AA \ spectral breaks indicative of young stellar
populations, and low stellar masses, resulting in high specific star
formation rates \citep{Rojas05}.

However, red quiescent galaxies do exist in voids, just with a lower
space density than blue, star forming galaxies
\citep{Croton05}. \citet{Croton08} show that the observed luminosity
function of void galaxies can be replicated with a \lcdm N-body
simulation and simple semi-analytic prescriptions for galaxy
evolution.  They explain the existence of red galaxies in voids as
residing in the few massive dark matter halos that exist in voids.
Their model requires some form of star formation quenching in massive
halos ($>\sim10^{12}$ \msun), but no additional physics that operates
only at low density needs to be included in their model to match the
data.  It is therefore the shift in the halo mass function in voids
that leads to different galaxy properties, not a change in the galaxy
evolution physics in low density environments.

\begin{figure}
\begin{center}
\scalebox{0.45}{\includegraphics{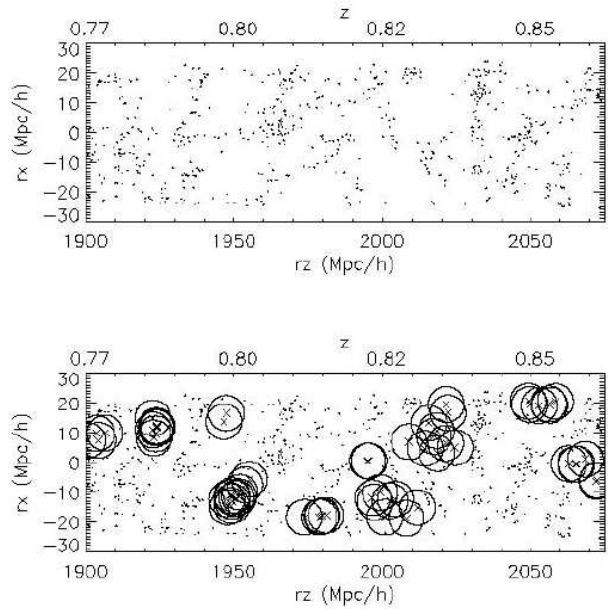}}
\end{center}
\caption{A schematic of the void probability function (VPF).  The top panel 
shows the comoving distribution of galaxies in a small portion of the 
DEEP2 survey (projected through 10 \mpch), while the lower panel shows 
a fraction of the empty spheres identified with a radius of 6 \mpch \
in the same volume (from Conroy et al. 2005).  Because the figure is projected
through one dimension, it may appear that galaxies reside inside of identified
voids; in three dimensions the voids contain no galaxies.
}
\end{figure}

\subsubsection{Void Probability Function}

In addition to identifying individual voids and the galaxies in them, one 
can study the statistical distribution of voids using the void probability
function (VPF).  Defined by \citet{White79}, the VPF is the probability that
a randomly placed sphere of radius $R$ within a point distribution 
will not contain any points (i.e., galaxies, see Figure 11).  The VPF is defined such that it
depends on the space density of points; therefore one must be careful when
comparing datasets and simulation results to ensure that the same number
density is used.  The VPF traces clustering in the weakly non-linear regime, 
not in the highly non-linear regime of galaxy groups and clusters.

\citet{Benson03} predict using \lcdm simulations that the VPF of galaxies 
should be higher than that of dark matter, that voids as traced by galaxies 
are much larger than voids traced by dark matter.  This results from the 
bias of galaxies compared to dark matter in voids and the fact that in this 
model the few dark matter halos that do exist in voids are low mass and 
therefore often do not contain bright galaxies.  \citet{Croton04} measure 
the VPF in the 2dFGRS dataset and find that it follows hierarchical scaling
laws, in that all higher-order correlation functions can be expressed in terms
of the two-point correlation function.  They find that even on scales of
$\sim$30 \mpch, higher-order correlations have an impact, and that the VPF of
galaxies is observed to be different than that of dark matter in simulations.

\citet{Conroy05} measure the VPF in SDSS galaxies at $z\sim0.1$ and DEEP2
galaxies at $z\sim1$ and find that voids traced by redder and/or brighter 
galaxy populations are larger than voids traced by bluer and/or fainter 
galaxies.  They also find that voids are larger in comoving coordinates at 
$z\sim0.1$ than at $z\sim1$; i.e., voids grow over time, as expected.  They show that 
the differences observed in the VPF as traced by different galaxy populations 
are 
entirely consistent with 
differences observed in the two-point correlation 
function and space density of these galaxy populations.  This implies that 
there does not appear to be additional higher-order information in voids 
than in the two-point function alone.  They also find excellent agreement
with predictions from \lcdm simulations that include 
semi-analytic models of galaxy evolution.

\citet{Tinker08} interpret the observed VPF in galaxy surveys in terms of 
the halo model (see Section 8 above).  
They compare the observed VPF in 2dFGRS and SDSS to halo model predictions 
constrained to match the two-point correlation function and number density 
of galaxies, using a model in which the dark matter halo occupation 
depends on mass only.  They find that with this model they can match the 
observed data very well, implying that there is no need for the 
suppression of galaxy formation in voids; i.e., galaxy formation does not
proceed differently in low-density regions.  They find that the sizes 
and emptiness of voids show excellent agreement with predictions of \lcdm
models for galaxies at low redshift to luminosities of $L\sim0.2 L^*$.

\subsection{Filaments}

Galaxy filaments -- long strings of galaxies -- 
are the largest systems seen in maps of large scale 
structure, and as such provide a key test of theories of structure formation. 
\citet{Aragon10} show that in simulations while filaments occupy only 
$\sim10$\% of the volume of space, they account for $\sim$40\% of the mass 
content of the $z=0$ Universe.
Measuring the typical and maximal length of filaments, as well as their 
thickness and average density, therefore constrains theoretical models. 
Various statistical methods have been proposed to identify and characterize 
the morphologies and properties of filaments (e.g. \citet{Sousbie08} and 
references therein).

In terms of their sizes, the largest length scale at which filaments are 
statistically significant, and hence identified as real objects, is 
50-80 \mpch, 
according to an analysis of galaxies in the Las Campanas Redshift Survey 
\citep[LCRS; ][]{Shectman96} by \citet{Bharadwaj04}.
They show that while there appear to be filaments in the survey on
longer scales, these arise from chance alignments and projection
effects and are not real structures.
\citet{Sousbie08} identify and study the length of filaments in SDSS, by 
identifying ridges in the galaxy distribution using the Hessian matrix 
($\partial^2\rho / \partial x_i \partial x_j$) and its eigenvalues (see Fig.~12). 
They find excellent 
agreement between observations and \lcdm numerical predictions for 
a flat, low $\Omega_{\rm matter}$ Universe.  They argue that filament 
measurements are not highly sensitive to observational effects such as 
redshift space distortions, edge effects, incompleteness or galaxy bias, 
which makes them a robust test of theoretical models.

\begin{figure}
\begin{center}
\scalebox{0.5}{\includegraphics{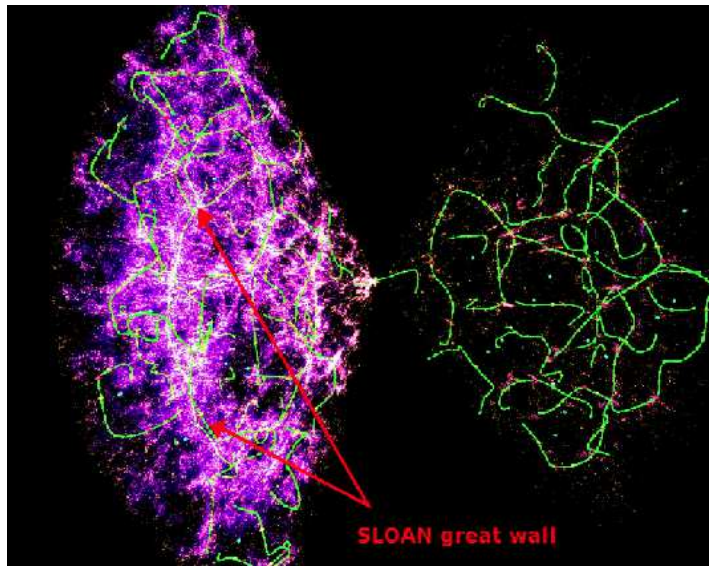}}
\end{center}
\caption{Filaments identified in the SDSS galaxy distribution 
(from Sousbie et al. 2008).  Individual filaments are shown in green 
overlaid on the galaxy density field show in purple.  The Sloan Great Wall 
is identified in the foreground, lying between the red arrows.
}
\end{figure}

\citet{Bond10b} use the eigenvectors of the Hessian matrix of the smoothed
galaxy distribution to identify filaments in both SDSS data and \lcdm 
simulations and find that the distribution of filaments lengths is roughly
exponential, with many more filaments of length $\lesssim$10 \mpch \ than 
$>20$ \mpch.  They find that the filament width distribution agrees between 
the SDSS data and N-body simulations.  The mean filament width depends on
the smoothing length; for smoothing scales of 10 and \mpch, the mean filament
widths are 5.5 and 8.4 \mpch. In \lcdm simulations they find that the 
filamentary structure in the dark matter density distribution 
is in place by $z=3$, tracing a similar pattern of density ridges.  This is in
contrast to what is found for voids, which become much more prominent and 
low-density at later cosmic epochs.  

\citet{Choi10} use the methods of \citet{Bond10b} to study the 
evolution of filamentary structure from $z\sim0.8$ to 
$z\sim0.1$ using galaxies from the DEEP2 survey and the SDSS.  
They find that neither the space density of filaments nor the distribution of 
filament lengths has changed 
significantly over the last seven Gyr of cosmic time, in agreement
with \lcdm numerical predictions.  The distribution of filament
widths has changed, however, in that the distribution is broader at lower
redshift and has a smaller typical width.  This observed evolution in
the filament width distribution naturally results 
from non-linear growth of structure and is consistent with the results on voids
discussed above, in that over time voids grow larger while filaments become 
tighter (i.e. have a smaller typical width) though not necessarily longer.

\section{Summary and Future}

This overview of our current understanding of the large-scale structure 
of the Universe has shown that quantitative measurements of
the clustering and spatial distribution of galaxies have wide applications 
and implications.  
The non-uniform structure reveals properties of both the galaxies and
the dark matter halos that comprise this large-scale structure.
Statistics such as the two-point correlation function 
can be used not only to constrain cosmological parameters but also to 
understand galaxy formation and evolution processes.  The advent of 
extremely large redshift surveys with samples of hundreds of thousands 
of galaxies has led to very precise measurements of the clustering of
galaxies at $z\sim0.1$ as a function of various galaxy properties such 
as luminosity, color, and stellar mass, influencing our understanding
of how galaxies form and evolve.  Initial studies at 
higher redshift have revealed that many of the general correlations that 
are observed between galaxy properties and clustering at $z\sim0$ were 
in place when the Universe was a fraction of its current age.  
As larger redshift surveys are carried out at 
higher redshifts, much more can be learned about how galaxy populations change 
with time.  
Theoretical interpretations of galaxy clustering measurements such 
as the halo occupation distribution model have also
recently made great strides in terms of statistically linking 
various properties of galaxies with those of their host dark matter halos. 
Such studies reveal not only how light traces mass on large scales but how 
baryonic mass and dark matter co-evolve with cosmic time.

There are many exciting future directions for studies of galaxy 
clustering and large-scale structure.  Precise cosmological 
constraints can be obtained using baryon acoustic oscillation signatures
observed in clustering measurements from wide-area surveys \citep{Eisenstein05}.
Specific galaxy populations can be understood in greater detail by 
comparing their clustering properties with those of galaxies in general.
For example, the clustering of different types of active galactic nuclei (AGN) 
can be used to constrain the AGN fueling mechanisms, lifetimes, and host galaxy 
populations \citep{Coil09}.  
As discussed above, measurements of galaxy clustering have the power to 
place strong constraints on contemporary models of galaxy formation and 
evolution
and advance our understanding of how galaxies populate and evolve within 
dark matter halos.

\

The author thanks James Aird, Mirko Krumpe, and Stephen Smith for providing
comments on earlier drafts of the text.


\end{document}